\documentclass[vecphys]{svmult}

\usepackage{makeidx}         
\usepackage{graphicx}        
\usepackage{multicol}        
\usepackage[bottom]{footmisc}
\usepackage{url}
\usepackage{amsfonts}
\usepackage{amssymb}
\usepackage{amsbsy}
\usepackage{epsfig}

\makeindex             

\newcommand{\BEQ}{\begin{equation}}     
\newcommand{\BEA}{\begin{eqnarray}}
\newcommand{\BD}{\begin{displaymath}}
\newcommand{\EEQ}{\end{equation}}       
\newcommand{\EEA}{\end{eqnarray}}
\newcommand{\ED}{\end{displaymath}}
\newcommand{\eps}{\varepsilon}          

\newcommand{\wit}[1]{\widetilde{#1}}    

\renewcommand{\vec}[1]{\boldsymbol{#1}} 

\begin{document}

\title*{Local scale-invariance in disordered systems}
\titlerunning{LSI in disordered systems} 
\author{Malte Henkel\inst{1} \and Michel Pleimling\inst{2}}
\institute{$^1$Laboratoire de Physique des Mat\'eriaux, 
Universit\'e Henri Poincar\'e Nancy I, B.P. 239, 
F - 54506 Vand{\oe}uvre-l\`es-Nancy Cedex, France\\
$^2$Department of Physics, 
Virginia Polytechnic Institute and State University,\\
Blacksburg, Virginia 24061-0435, USA}

%
%
\maketitle
\input{epsf.sty} 

\abstract{Dynamical scaling and ageing in disordered systems far from
equilibrium is reviewed. 
Particular attention is devoted to the question to what extent a recently
introduced generalization of dynamical scaling to local scale-invariance can 
describe data for either non-glassy systems quenched to below $T_c$ or else
for spin glasses at criticality. 
The dependence of the scaling behaviour on the 
distribution of the random couplings is discussed.
It is shown that finite-time corrections to scaling can
become quite sizable in these systems.
Numerically determined ageing quantities are confronted with
available experimental results.
}\\

%
%

\section{Introduction}

Understanding cooperative phenomena far from equilibrium poses one of the
most challenging research problems of present-day many-body physics. 
At the same time, the practical handling of many of these materials has
been pushed to great sophistication, and a lot of 
practical knowledge about them exists since prehistoric times. 
Paradigmatic examples of such system are glasses. 
In many cases, they are made by rapidly cooling 
(`quenching') a molten liquid
to below some characteristic temperature-threshold. If this cooling happens
rapidly enough, normal crystallization no longer takes place and the material 
remains in some non-equilibrium state. These non-equilibrium states may
at first and even second sight look very stationary -- everyone has probably
seen in archaeological museums intact specimens of Roman glass or even older
tools from the Paleolithic or old-stone-age -- after all, 
obsidian or fire-stone is a quenched volcanic melt. But since the material 
is not at equilibrium, at least in principle it is possible 
(and it does happen very often in practice) that over time the 
properties of the material change - in other words, the material 
ages.\footnote{Recall that {\em physical ageing} as it is understood here 
comes from reversible microscopic processes, whereas chemical or biological 
ageing may come from the action of essentially irreversible (bio-)chemical 
processes.} 
The properties of such non-equilibrium systems depend on the time 
-- their {\em age} -- since they were brought out of equilibrium and this
is colloquially referred to as {\em ageing behaviour}.

Although the first systematic studies of ageing were carried out in the by now
classic experiments of Struik \cite{Stru78} in polymeric glasses,  
it has been realized in recent years that very similar phenomena
already occur in non-disordered and non-frustrated systems which hence are also
commonly referred to as {\em ageing}. In order to present the essence of the 
ageing phenomenon, we begin by considering 
a simple Ising model, made from spin 
variables $\sigma_i=\pm 1$ attached to
each site $i$ of a hypercubic lattice and an interaction as given by the 
classical hamiltonian with the usual nearest-neighbour interactions 
\BEQ
{\cal H}=-J \sum_{(i,j)} \sigma_i \sigma_j
\EEQ
where $J>0$ is the exchange integral and the sum extends over pairs of nearest 
neighbours. In $d>1$ dimensions, this system has a phase-transition at some
critical temperature $T_c>0$ which separates a disordered phase at high
temperatures with a single thermodynamically stable state from an ordered
phase at low temperatures with two thermodynamically stable equilibrium states.
It is well-established that such simple
Ising models may be used to discuss uniaxial magnets, binary alloys 
or even liquid crystals. The motion of the spins is generated by coupling the 
model to a thermal bath of temperature $T$. 
A possible way of realizing this is through the
so-called heat-bath dynamics which is defined 
by the stochastic rule
\begin{equation}
\sigma_i(t+\Delta t) = \pm 1 \;\;\;
\mbox{\rm with probability $\left[ 1\pm \tanh( h_i(t)/T)\right]/2$}
\end{equation}
where $\Delta t$ is the time increment, the local time-dependent 
field is $h_i(t)=\sum_{y(i)} \sigma_{y(i)}(t)$ and
$y(i)$ runs over the nearest neighbours of the site $i$. 
It is well-known that
this rule satisfies detailed balance and hence the system 
evolves towards the equilibrium probability distribution 
$P_{\rm eq}=Z^{-1} \exp(-{\cal H}/T)$, where
$Z$ is the canonical partition function \cite{vKam92,Nara01,Zwan01}. 
The system is prepared at some initial temperature $T_{\rm ini}$ far above the
critical temperature $T_c>0$. The initial time $t=0$ is defined by coupling the
system to the thermal bath at some low temperature $T<T_c$ and starting the
dynamics. During the simulation, the temperature $T$ is kept fixed and one
observes the time-dependence of observables such as correlation functions or
susceptibilities.\footnote{The chosen dynamics is such that the total 
average magnetization 
$M(t) = \langle\sigma(t)\rangle = \sum_{i}\langle\sigma_i(t)\rangle$ remains 
at its initial value $M(0)=0$. Throughout, we shall always use this initial
condition unless explicitly stated otherwise.} 

\begin{figure}[t]
\centerline{\epsfxsize=3.5in\epsfbox
{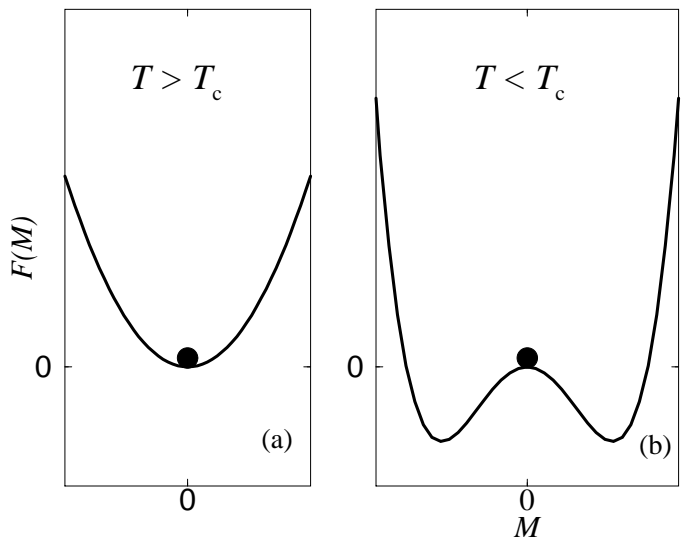}
}
\caption{Free energy $F(M)$ as a function of the magnetization $M$ 
of a simple ferromagnet at (a) an initial high temperature
$T>T_c$ before the quench and (b) after the quench to a low temperature $T<T_c$.
\label{LNPHePl:Abb1}}
\end{figure}

Qualitatively, the behaviour of the system
can be illustrated through the equilibrium free energies at the temperatures
$T_{\rm ini}$ and $T$, see figure~\ref{LNPHePl:Abb1}. 
Before the quench, the system is at equilibrium with respect to the
initial temperature $T_{\rm ini}\gg T_c$ and sits at the minimum of the free
energy, as indicated by the black ball in figure~\ref{LNPHePl:Abb1}a. 
If one would perturbe this system slightly, it would relax rapidly,
i.e. with a
finite relaxation time $0<\tau<\infty$, towards this unique equilibrium state.
On the other hand, immediately
after the quench the system does not yet have had the time to evolve but, with
respect to the new equilibrium, its free energy is no longer minimal, see
figure~\ref{LNPHePl:Abb1}b. Rather,
two new local minima of the free energy appear, which correspond to the two
equivalent ordered states of the system. Because 
of the competition between these
two equivalent equilibrium states, the system as a whole cannot relax rapidly 
to one of them but rather undergoes a slow dynamics, 
with formally infinite relaxation times. Locally, each 
spin will be subject to the time-dependent field
$h_i(t)$ coming from its neighbours and this field will 
tilt the balance between the two equivalent equilibrium states of 
figure~\ref{LNPHePl:Abb1}b in favour of 
one or the other. Physically, this means that the system will rapidly decompose
into ordered domains and the slow long-time dynamics of this 
domain growth will be determined by the
motion of the domain walls between these ordered domains. 
This slow (non-exponential) dynamics
is the {\em first defining property} of ageing systems. 

\begin{figure}
\centerline{\psfig{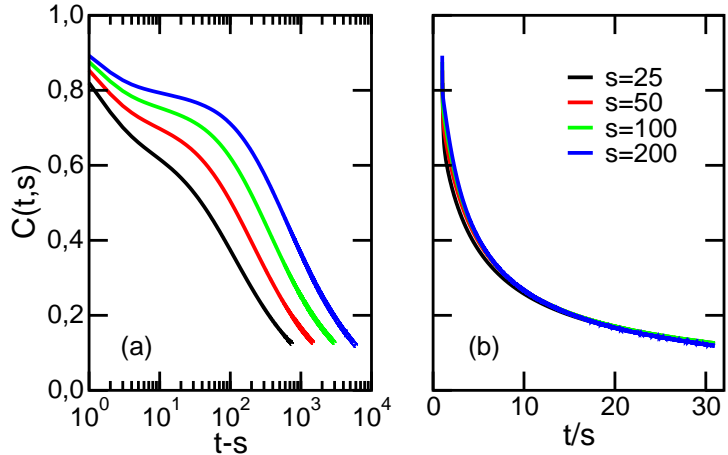}}
\caption[Scaling of three-dimensional Ising model]{(a) Ageing and (b) dynamical 
scaling of the two-time autocorrelation
function $C(t,s)$ in the three-dimensional Glauber-Ising model quenched to 
$T=3<T_c$, for several values of the waiting time $s$. 
\label{LNPHePl:Abb2}
}
\end{figure} 

Another aspect of this non-equilibrium dynamics (since in a spatially infinite
system none of the equilibrium states will be reached in a finite time) becomes
apparent if one considers a quantity like the two-time autocorrelation function
of spins at site $i$ at times $t$ and $s$ 
\begin{equation}
C(t,s) = \left\langle \sigma_i(t) \sigma_i(s) \right\rangle
\end{equation}
which by spatial translation-invariance is independent of the chosen site $i$. 
In figure~\ref{LNPHePl:Abb2}a data for $C(t,s)$ plotted
over against the time difference $t-s$ are displayed for the 
three-dimensional Ising model. We see that the data depend on {\em both} 
$t-s$ and $s$, hence time-translation invariance is broken
and the system ages. Further,  with increasing 
values of the waiting time $s$, the system becomes `stiffer' and a plateau
close to the equilibrium value $C_{\rm eq}=M_{\rm eq}^2$ develops when
$t-s$ is not too large before the correlations fall off rapidly when
$t-s\to\infty$. Together 
with the slow dynamics mentioned above, this breaking of time-translation
invariance is the {\em second defining property} of ageing systems. 
In principle, this could
mean that the details of the dynamics of ageing systems might depend on the
entire prehistory of the sample under study, which would make any attempt to
formulate a general theory for such systems hopeless. However, 
a great simplification, due to dynamical scaling, 
is apparent in figure~\ref{LNPHePl:Abb2}b where the {\em same}
data for $C(t,s)$, when plotted over against $t/s$, neatly collapse onto 
a single
curve, if only the time $s$ is large enough and also $t-s$ is not too small. 
Since in domain coarsening of simple magnets one expects that the linear 
size of the ordered domains is $L=L(t)\sim t^{1/z}$ when $t$
is large enough and $z$ is the dynamical exponent \cite{Bray94}, 
the collapse in figure~\ref{LNPHePl:Abb2}b means that $C(t,s) = f(L(t)/L(s))$, 
or in other words
$L(t)$ is the only relevant length-scale at time $t$. Dynamic scaling 
(although more general forms for $L(t)$ than simple power laws are of course
possible) is the {\em third essential property} of ageing systems. 

These three basic properties of ageing  
systems are also found in glassy systems. 
An important property of glasses is the presence of frustrations which prevent
the relaxation of all local degrees of freedom. In consequence, 
the free-energy
landscape of glasses can be very complex, with many local minima. The classic
example for ageing behaviour was observed by Struik \cite{Stru78} in 
studying the mechanical 
properties of polymeric glasses which after a quench from the molten phase
to low temperatures (i) relax very slowly (typical time-scale of years), 
(ii) show clear evidence of the
breaking of time-translation invariance and furthermore, (iii) 
the experimental data for the time-dependent creep curves of the 
mechanical response can all be mapped onto a single master curve. 
Remarkably, that master curve turned out to be {\em the same} for materials
as different as polymers such as PVC or PMMA, sugar or even metals like lead~! 
Evidently, there are {\em universal} scaling functions in 
ageing which exactly because
of their universality one may hope to be able to understand theoretically. 
Returning to simple magnets, that universality in the kinetics of coarsening
(with a non-conserved order-parameter) after a quench to $T<T_c$ 
is captured through the celebrated 
Allen-Cahn equation \cite{Alle79} which states that
the velocity $v$ of the domain walls which separate the ordered domains is 
related to the curvature $K$ in $d$ dimensions via the purely 
geometric relationship $v = (d-1) K$, quite 
independently of any details of the interactions of the spins.

\begin{figure}[t]
\centerline{
\epsfysize=95mm
\epsffile{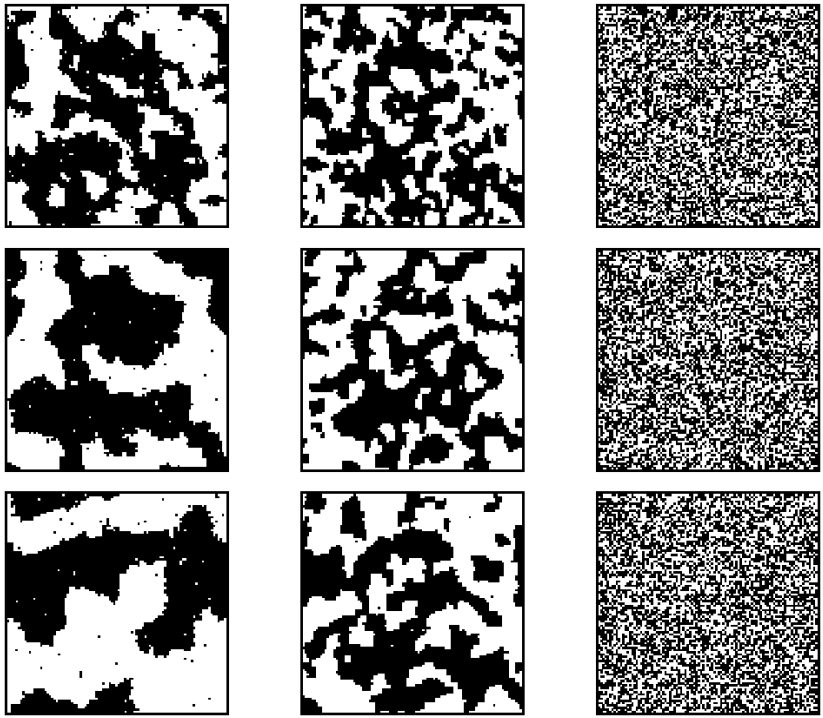} }
\caption[Growth of clusters]{Snapshots illustrating the growth
of clusters in three variants of the kinetic Ising model, at times $t=25, 100$ 
and $225$ after the quench, from top to bottom. 
The left column shows the two-dimensional 
Glauber-Ising model without disorder quenched to $T=1.5<T_c$ and the middle 
column shows the random bond Ising model in two dimensions with 
$\varepsilon=2$ and $T=0.7<T_c$. 
The right column shows a section of the three-dimensional Ising spin glass with 
binary disorder at $T=0.8<T_c$. 
\label{LNPHePl:amas3fois}
}
\end{figure} 

The microscopic evolution which underlies the above statements about the
properties of macroscopic observables is illustrated in
figure~\ref{LNPHePl:amas3fois}. In the first column, snapshots of 
configurations
of the two-dimensional Glauber-Ising model without disorder are shown for three 
different times after the quench from a fully disordered initial configuration 
into the ordered phase. Clearly, ordered 
domains form very rapidly and continue
to grow up to the system size. In order to appreciate the effects of disorder,
we next consider a random-bond Ising model defined by the classical 
hamiltonian
\BEQ \label{gl:disI}
{\cal H}_{\rm dis} = - \sum_{(i,j)} J_{i,j} \sigma_i \sigma_j
\EEQ
where the $J_{i,j}$ are random variables uniformly distributed over the
interval $[1-\varepsilon/2, 1+\varepsilon/2]$ with $0 < \varepsilon \leq 2$. Snapshots from this 
two-dimensional model
are shown for $\varepsilon=2$ and $T=0.7<T_c$ 
in the middle column of figure~\ref{LNPHePl:amas3fois}. Again, 
detectable ordered domains from rapidly but with respect to the non-disordered
Glauber-Ising model, although the clusters look broadly similar, 
one sees that the growth of the clusters proceeds much more
slowly. Finally, we consider an Ising spin glass, which can again be defined
by the hamiltonian (\ref{gl:disI}) but now the random couplings $J_{i,j}=\pm 1$
can take both positive and negative values with equal probabilites. 
The right column in figure~\ref{LNPHePl:amas3fois} illustrates the evolution of 
the three-dimensional Ising 
spin glass with binary disorder ($J_{i,j}=\pm 1$ with equal probability) at the 
temperature $T=0.8< T_c$. It is clear that the
presence of both ferromagnetic and antiferromagnetic interactions no longer
permits the formation of ordered clusters and with respect to the two other
cases, the evolution appears considerably slower.  
We shall formalize this later on when we discuss the growth laws of the 
time-dependent characteristic length.

Universality as seen in experiments \cite{Stru78} or manifest in the 
Allen-Cahn equation may signal the presence of deeper dynamical symmetries. 
In this review, we shall describe the available evidence which points
to the possibility that the well-established dynamical scaling found
in ageing systems may be included into a larger group of 
{\em local} scale-transformations. As a necessary preparation we shall 
begin by briefly recalling the main features of dynamical scaling. 
The topic has been under intensive study, see
\cite{Bray94,Cate00,Cugl02,Godr02,Cris03,Kawa04,Cala05,Gamb06,Henk06,Henk07a} 
for reviews. 

It is useful to discuss the properties of the coarse-grained order-parameter,
denoted by $\phi(t,\vec{r})$ at time $t$ at the location $\vec{r}$. Most of
our discussion is for the case that the order-parameter is {\em non-conserved} 
by the dynamics and that the initial
state is totally disordered with $\langle \phi(0,\vec{r})\rangle=0$, 
unless explicitly stated otherwise. 
It has become common to study ageing behaviour through 
the two-time autocorrelation and (linear) autoresponse functions
\BEA
C(t,s) &=& \langle \phi(t,\vec{r}) \phi(s,\vec{r}) \rangle ~~ 
\sim s^{-b} f_C(t/s) 
\label{gl:sC} \\
R(t,s) &=& \left.\frac{\delta\langle\phi(t,\vec{r})\rangle}{\delta 
h(s,\vec{r})}\right|_{h=0}
\sim s^{-1-a} f_{R}(t/s)
\label{gl:sR}
\EEA
where $h(s,\vec{r})$ is the magnetic field conjugate to $\phi$ at time $s$ 
and location $\vec{r}$. 
The scaling behaviour is expected to apply
in the so-called {\em ageing regime} where 
\BEQ
t,s\gg t_{\rm micro} 
\;\; \mbox{\rm and} \;\; t-s\gg t_{\rm micro}
\EEQ
where $t_{\rm micro}$ is a microscopic time-scale. We emphasize the 
importance of the third condition for the validity of dynamical scaling. 
This makes the above qualitative observations drawn from 
figure~\ref{LNPHePl:Abb2} more precise. In particular, we see in 
figure~\ref{LNPHePl:Abb2} that for {\em finite} values of $s$ scaling is not
perfect: indeed corrections to dynamical scaling are detectable when 
$t/s\approx 1$. 

The distance of such systems from a global equilibrium state can be 
measured through the fluctuation-dissipation ratio, defined as \cite{Cugl94b}
\BEQ \label{gl:rfd}
X(t,s) := T R(t,s) \left( \frac{\partial C(t,s)}{\partial s}\right)^{-1}.
\EEQ
At equilibrium, $X(t,s)=1$ from the fluctuation-dissipation theorem. One often
considers the limit fluctuation-dissipation ratio\footnote{The order of the 
limits is crucial, since 
$\lim_{t\to\infty}\left(\lim_{s\to\infty} X(t,s)\right)=1$.}  
\BEQ
X_{\infty} := \lim_{s\to\infty} \left( \lim_{t\to\infty} X(t,s)\right)
= \lim_{y\to \infty} \left( \lim_{t, s\to\infty} 
\bigl. X(t,s)\bigr|_{y=t/s} \right).
\EEQ
For quenches to below $T_c$, one usually has $X_{\infty}=0$ but for
{\em critical} quenches onto $T=T_c$, it has been proposed by 
Godr\`eche and Luck that $X_{\infty}$ should be a {\em universal} 
number \cite{Godr00b}, since it can be written as a ratio of two scaling 
amplitudes. The universality of $X_{\infty}$ has been thoroughly confirmed for 
systems relaxing towards equilibrium steady-states, and can be 
theoretically validated in a field-theoretical setting, 
see \cite{Cris03,Cala05} for recent reviews. On the other hand, we mention
here that quite similar scaling behaviour is also found for systems containing
irreversible chemical reactions and hence relaxing towards non-equilibrium
steady-states. In that case, the definitions of $X(t,s)$ and of $X_{\infty}$ 
have to be reconsidered, see \cite{Henk07a} for a recent review.  

Furthermore, in writing eqs.~(\ref{gl:sC},\ref{gl:sR}) 
it was tacitly assumed that the scaling derives from the 
algebraic time-dependence of a single characteristic length-scale growing
algebraically with time 
$L(t)\sim t^{1/z}$ which measures the linear size of correlated or ordered 
clusters and where $z$ is the dynamic exponent. That is indeed the
case for  simple magnets whose kinetics may be described in terms
of a simple Ising model, e.g. with Glauber-type dynamics, where it can even
be shown that $z=2$ \cite{Bray94,Bray94b}.  
Then the above forms (\ref{gl:sC},\ref{gl:sR}) define the 
non-equilibrium exponents $a$ and $b$ and the scaling functions $f_C(y)$ and
$f_R(y)$. For large arguments $y\to \infty$, one generically expects
\BEQ
f_C(y) \sim y^{-\lambda_C/z} \;\; , \;\;
f_R(y) \sim y^{-\lambda_R/z}
\EEQ
where $\lambda_C$ and $\lambda_R$, respectively, are known as autocorrelation
\cite{Fish88,Huse89} and autoresponse exponents \cite{Pico02}. 

\begin{table}[b]
\caption{Values of the non-equilibrium exponents $a$, $b$ and $z$ 
for non-conserved ferromagnets with $T_c>0$ and a vanishing initial 
magnetization. The non-trivial critical-point value $z_c$ is 
model-dependent.\label{TabExp1}}
\renewcommand{\arraystretch}{1.4}
\setlength\tabcolsep{10pt}
\begin{center}\begin{tabular}{|l|lll|l|}  
\hline\hline\noalign{\smallskip}
        & $a$            & $b$            & $z$   & Class \\ 
\noalign{\smallskip}\hline\hline\noalign{\smallskip}
$T=T_c$ & $(d-2+\eta)/z$ & $(d-2+\eta)/z$ & $z_c$ & L \\ \hline
$T<T_c$ & $(d-2+\eta)/z$ & 0              & 2     & L \\
        & $1/z$          & 0              & 2     & S \\ \hline\hline
\end{tabular}\end{center}
\end{table}

While in non-disordered magnets with short-ranged initial conditions one usually
has $\lambda_C=\lambda_R$, this is not necessarily so if either
of these conditions is relaxed. From a field-theoretical point of view it 
is known that for a non-conserved order-parameter 
the calculation of $\lambda_{C,R}$ requires an independent renormalization 
and hence one cannot expect to find a scaling relation between
these and equilibrium exponents (including $z$) \cite{Jans89}. 
On the other hand, the values of the exponents $a$ and $b$ are known. For
quenches to $T=T_c$, the relevant length-scale is set by the time-dependent
correlation length $L(t)\sim \xi(t) \sim t^{1/z}$ and this leads to
$a=b=(d-2+\eta)/z$, where $\eta$ is a standard equilibrium exponent. For 
quenches into the ordered phase $T<T_c$, one usually observes simple scaling
of $C(t,s)=f_C(t/s)$, hence $b=0$.\footnote{This needs no longer be the case 
when the ageing close to a free surface is considered \cite{Baum06a}.} 
The value of $a$ depends on whether the equilibrium
correlator is short- or long-ranged, respectively. These may be referred to 
as classes S and L, respectively, and 
one has, see e.g. \cite{Cate00,Henk02a,Henk03e}
\BEQ
\hspace{-1truecm} 
C_{\rm eq}(\vec{r})\sim \left\{\begin{array}{l} 
e^{-|\vec{r}|/\xi} \\
|\vec{r}|^{-(d-2+\eta)} \end{array} \right. 
\;\; \Longrightarrow \;\; 
\left\{\begin{array}{l} \mbox{\rm class S} \\ 
\mbox{\rm class L}\end{array}\right.
\;\; \Longrightarrow \;\; 
a = \left\{ \begin{array}{c} 1/z \\ (d-2+\eta)/z \end{array} \right.
\EEQ
Examples for short-ranged models (class S) include the Ising or Potts 
models in $d>1$
dimensions (and $T<T_c$), while all systems quenched to criticality, or
the spherical model or the two-dimensional XY model below the 
Kosterlitz-Thouless transition are examples for long-ranged systems (class L). 

The values of these exponents are again collected in table~\ref{TabExp1}. They
have been extensively confirmed in many numerical studies, see
\cite{Henk02a,Henk03e,Chat03,Abri04a,Abri04b,Lore06}, and are reproduced in all 
known analytically solvable models. 

In section~2, we recall some of the main facts related to an extension of
dynamical scaling towards a local scale-invariance (LSI), first for a
dynamical exponent $z=2$ and later on possible extensions for $z\ne 2$ 
and review its application to non-disordered systems. In section~3 we discuss
the scaling of responses and correlators in disordered, but non-glassy systems
and shall present the available evidence that LSI should be extendable to
this class of systems. In section~4 we consider the critical Ising spin glass,
present evidence that its non-equilibrium properties depend on the
distribution of the coupling constants and show to what extent the data
can be explained in terms of LSI. We summarize in section~5. 

\section{Local scale-invariance without disorder}

In equilibrium critical phenomena, it is well-known that the standard 
scale-invariance can, under quite weak conditions, be extended to
a {\em conformal} invariance. Roughly, a conformal transformation is
a scale-transformation $\vec{r}\mapsto b \vec{r}$ with a space-dependent 
rescaling factor $b=b(\vec{r})$ (such that angles are kept unchanged).  
In particular, in two dimensions conformal invariance allows to derive
from the representation theory of the conformal (Virasoro) algebra the 
possible values of the critical exponents, to set up a list of possible 
universality classes, to calculate explicitly all $n$-point correlation 
functions and so on, see \cite{Bela84,Card90,Fran97,Henk99}. 
One might wonder whether a similar extension might be possible 
at least in some instances of dynamical scaling and further ask {\it whether
response functions or correlation functions might be found from their
covariance under some generalized dynamical scaling with a space-time-dependent
rescaling factor} $b=b(t,\vec{r})$ \cite{Henk94,Henk02}~?  

We shall first consider the case of phase-ordering where $L(t)\sim t^{1/z}$
with $z=2$ is 
known \cite{Bray94} and later describe how this might be generalized to more
general values of $z\ne 2$. In particular we enquire what
can be said about the scaling functions $f_{C,R}(y)$ 
in a model-independent way. 

A useful starting point is to consider the symmetries of the free diffusion 
(or free Schr\"odinger) equation
\BEQ \label{gl:diffu}
2{\cal M} \partial_t \phi = \Delta \phi
\EEQ
where $\Delta=\vec{\nabla}\cdot\vec{\nabla}$ is the spatial laplacian and 
the `mass' $\cal M$ plays the r\^ole of a kinetic coefficient. Indeed, 
it was already shown by Lie more than a century ago that this equation
has more symmetries than the obvious translation- and rotation-invariances. 
Consider the so-called {\em Schr\"odinger-group} defined through the
space-time transformations
\BEQ \label{gl:5:2:Schr}
t \mapsto t' = \frac{\alpha t+\beta}{\gamma t+\delta} \;\; ; \;\;
\vec{r} \mapsto \vec{r}' = 
\frac{\matrix{R}\vec{r} + \vec{v}t + \vec{a}}{\gamma t+\delta} 
\;\; , \;\;
\alpha\delta - \beta\gamma =1
\EEQ
where $\alpha,\beta,\gamma,\delta,\vec{v},\vec{a}$ are real (vector) 
parameters and $\matrix{R}$ is a rotation matrix in $d$ spatial dimensions. 
The group acts projectively on a solution $\phi$ of the diffusion 
equation through 
$(t,\vec{r})\mapsto g(t,\vec{r})$, $\phi\mapsto T_g \phi$
\BEQ \label{gl:5:2:Schrpsi}
\left(T_g \phi\right)(t,\vec{r}) = 
f_g(g^{-1}(t,\vec{r}))\,\phi(g^{-1}(t,\vec{r}))
\EEQ
where $g$ is an element of the Schr\"odinger group and the companion 
function reads \cite{Nied72,Perr77}
\BEQ \label{gl:5:2:Schrf}
f_{g}(t,\vec{r}) = (\gamma t+\delta)^{-d/2} 
\exp\left[ -\frac{{\cal M}}{2} \frac{\gamma \vec{r}^2+
2\matrix{R}\vec{r}\cdot(\gamma\vec{a}-\delta\vec{v})+\gamma\vec{a}^2-
t\delta\vec{v}^2+2\gamma\vec{a}\cdot\vec{v}}{\gamma t+\delta}\right].
\EEQ
It is then natural to include also arbitrary phase-shifts of the wave function 
$\phi$ within the Schr\"odinger group {\sl Sch}($d$). In what follows, 
we denote 
by $\mathfrak{sch}_d$ the Lie algebra of {\sl Sch}($d$). The Schr\"odinger
group so defined is the largest group which maps {\em any} solution of the
free Schr\"odinger equation (with $\cal M$ fixed) onto another solution. 
For a simple illustration, consider the case $d=1$ and define the 
Schr\"odinger operator
\BEQ
{\cal S} := 2 M_0 X_{-1} - Y_{-1/2}^2.
\EEQ
The Schr\"odinger Lie algebra
$\mathfrak{sch}_1=\langle X_{-1,0,1},Y_{-\frac{1}{2},\frac{1}{2}},M_0\rangle$
is spanned by the infinitesimal generators of temporal and spatial
translations ($X_{-1},Y_{-1/2}$), Galilei-transformations ($Y_{1/2}$),
phase shifts ($M_0$), space-time dilatations with $z=2$ ($X_0$) and so-called
special transformations ($X_1$). Explicitly, the generators read \cite{Henk94}
\BEA
X_n &=& -t^{n+1}\partial_t -\frac{n+1}{2} t^n r\partial_r -\frac{n(n+1)}{4}
{\cal M} t^{n-1} r^2 - \frac{x}{2}(n+1) t^n \nonumber \\
Y_m &=& -t^{m+1/2}\partial_r -\left( m+\frac{1}{2}\right) {\cal M} t^{m-1/2} r
\label{gl:5:2:SchrGen} \\
M_n &=& -{\cal M} t^n \nonumber
\EEA
Here $x$ is the scaling dimension and ${\cal M}$ is the {mass} of the
scaling operator $\phi$ on which these generators act. 
The non-vanishing commutation relations are
\BEA
\left[ X_n , X_{n'} \right] &=& (n-n') X_{n+n'} \;\; , \;\;
\left[ X_n , Y_m \right] \:=\: \left(\frac{n}{2}-m\right) Y_{n+m} 
\nonumber \\ 
\left[ X_n , M_{n'} \right] &=& -n' M_{n+n'} \;\; , \;\; 
\left[ Y_m , Y_{m'} \right] \:=\: (m-m') M_{m+m'} 
\label{gl:5:2:SchAlg}
\EEA
The invariance of the diffusion equation under the action of $\mathfrak{sch}_1$
is now seen from the following commutators which follow from the
explicit form (\ref{gl:5:2:SchrGen})
\BEA
\left[{\cal S}, X_{-1}\right] &=& \left[ {\cal S}, Y_{\pm 1/2} \right]
\:=\: \left[ {\cal S}, M_0\right] \:=\: 0 
\nonumber \\
\left[{\cal S}, X_0 \right] &=& -{\cal S}
\;\; , \;\; \qquad \;\;\:
\left[ {\cal S}, X_1 \right] \:=\: -2t {\cal S} - (2x-1) M_0 
\label{gl:LNP:Scomm}
\EEA
Therefore, {\em for any solution $\phi$ 
of the Schr\"odinger equation ${\cal S}\phi=0$ with
scaling dimension $x=1/2$, the infinitesimally transformed solution 
${\cal X}\phi$ with ${\cal X}\in\mathfrak{sch}_1$ also satisfies the 
Schr\"odinger equation ${\cal S}{\cal X}\phi=0$} \cite{Kast68,Nied72,Hage72}. 
For applications to ageing, we must consider to so-called {\em ageing algebra} 
$\mathfrak{age}_1 =\langle X_{0,1},Y_{-\frac{1}{2},\frac{1}{2}},M_0\rangle
\subset \mathfrak{sch}_1$ (without time-translations) 
which is a true subalgebra of $\mathfrak{sch}_1$. 
Extensions to $d>1$ are straightforward. 

What is the usefulness of knowing dynamical symmetries of free, simple
diffusion for the understanding of non-equilibrium kinetics ? 
One way of setting up the problem would be to write down a stochastic
Langevin equation for the order-parameter. The simplest case is usually
considered to be a dynamics without macroscopic conservation laws (model A), 
where one would have \cite{Hohe77} 
\BEQ \label{gl:5:Langevin}
2{\cal M} \frac{\partial\phi}{\partial t} = 
\Delta \phi - \frac{\delta {\cal V}[\phi]}{\delta \phi} +\eta
\EEQ
where $\cal V$ is the Ginzburg-Landau potential and $\eta$
is a gaussian noise which describes the coupling to an external heat-bath and
the initial distribution of $\phi$. At first sight, there appear to be no
non-trivial symmetries, since the noise term $\eta$ is incompatible with
a Galilei-invariant equation~(\ref{gl:5:Langevin}). 
To understand this physically, consider a magnet which
is at rest with respect to a homogeneous heat-bath at temperature $T$. 
If the magnet is moved with a constant velocity with respect to 
the heat-bath, the effective
temperature will now appear to be direction-dependent, and the heat-bath
is no longer homogeneous. However, that difficulty can be avoided as 
follows \cite{Pico04}: {\em split the Langevin equation into a
`deterministic' part with non-trivial symmetries and a `noise' part and then
show using these symmetries that all averages can be reduced exactly to
averages within the deterministic, noiseless theory}. Technically, one
first constructs in the standard fashion (Janssen-de Dominicis procedure)
\cite{deDo78,Jans92} the associated stochastic field-theory with action 
$J[\phi,\wit{\phi}]$ where $\wit{\phi}$ is the response field associated to 
the order-parameter $\phi$. Second, decompose the action into two parts
\BEQ \label{gl:5:JanDom}
J[\phi,\wit{\phi}] = J_0[\phi,\wit{\phi}] + J_b[\wit{\phi}]
\EEQ
where  
\BEQ \label{gl:5:JanDomdet}
J_0[\phi,\wit{\phi}]=\int_{\mathbb{R}_+\times\mathbb{R}^d}\!\D t\D\vec{r}\; 
\wit{\phi}\left(2{\cal M}\partial_t\phi-
\Delta\phi+\frac{\delta{\cal V}}{\delta\phi}\right)
\EEQ 
contains the terms coming from the `deterministic' part of
the Langevin equation ($\cal V$ is the self-interacting `potential') whereas
\BEQ \label{gl:5:JanDombruit}
\hspace{-1truecm}
J_b[\wit{\phi}] = -T \int_{\mathbb{R}_+\times\mathbb{R}^d}\!\D t\D\vec{r}\: 
\wit{\phi}(t,\vec{r})^2 
-\frac{1}{2}
\int_{\mathbb{R}^{2d}} \!\D\vec{r}\,\D\vec{r}'\:  
{\wit{\phi}}(0,\vec{r})a(\vec{r}-\vec{r}'){\wit{\phi}}(0,\vec{r}')
\EEQ
contains the terms coming either from the noisy coupling to the heat bath
or else from the average over the disordered initial conditions \cite{Jans92}. 
It was assumed here that $\langle \phi(0,\vec{r})\rangle =0$ and
$a(\vec{r})$ denotes the initial two-point correlator 
\BEQ
a(\vec{r}) := C(0,0;\vec{r}+\vec{r}',\vec{r}') = 
\langle \phi(0,\vec{r}+\vec{r}') \phi(0,\vec{r}')\rangle =a(-\vec{r})
\EEQ
while the last relation follows from spatial translation-invariance 
which we shall admit throughout. Averages are of course calculated from the
functional integral
\BEQ
\langle {\cal A}\rangle = \int \!{\cal D}\phi{\cal D}\wit{\phi}\: 
{\cal A}[\phi,\wit{\phi}] \, e^{-J[\phi,\wit{\phi}]}.
\EEQ

It is instructive to consider briefly the case of a free field, where
${\cal V}=0$. Variation of (\ref{gl:5:JanDom}) with respect to $\wit{\phi}$ and
$\phi$, respectively, then leads to the equations of motion
\BEQ \label{gl:5:equamovi}
2{\cal M}\partial_t \phi = \Delta \phi + T \wit{\phi} \;\; , \;\;
-2{\cal M}\partial_t\wit{\phi} = \Delta\wit{\phi}.
\EEQ
The first one of those might be viewed as a Langevin equation if $\wit{\phi}$ is
interpreted as a noise. Comparison of the two equations of motion 
(\ref{gl:5:equamovi}) shows that if the order-parameter $\phi$ 
is characterized by
the `mass'\index{mass} $\cal M$ (which by physical convention is positive), 
then the associated response field $\wit{\phi}$ is 
characterized by the {\em negative} mass $-{\cal M}$. 
This characterization remains valid beyond free fields. 

We now concentrate on actions $J_0[\phi,\wit{\phi}]$ which
are Galilei-invariant. This means that if $\langle .\rangle_0$ denotes the
averages calculated only with the action $J_0$, 
the Bargman superselection rules \cite{Barg54}
\BEQ \label{gl:5:Bargman}
\left\langle \underbrace{\phi\ldots\phi}_n \; \underbrace{\wit{\phi}\ldots
\wit{\phi}}_m \right\rangle_0 \sim \delta_{n,m}
\EEQ
hold true. It follows that both response and correlation functions can be
exactly expressed in terms of averages with respect to the deterministic part
alone. For example (we suppress
for notational simplicity the spatial coordinates) \cite{Pico04}
\BEQ
\hspace{-1truecm} 
R(t,s) = \left.\frac{\delta\langle \phi(t)\rangle}{\delta h(s)}\right|_{h=0}
= \left\langle \phi(t) \wit{\phi}(s) \right\rangle
= \left\langle \phi(t) \wit{\phi}(s)\, e^{-J_b[\wit{\phi}]} \right\rangle_0
= \left\langle \phi(t) \wit{\phi}(s) \right\rangle_0 
\label{gl:5:Rrauschlos}
\EEQ
where the `noise' part of the action was included in the
observable and the Bargman superselection rule
(\ref{gl:5:Bargman}) was used. In other words, 
{\em the two-time response function $R(t,s)=R_0(t,s)$ 
does not depend explicitly on the `noise' at all}~! The correlation function
is reduced similarly \cite{Pico04}
\BEA
\lefteqn{ 
C(t,s;\vec{r}) = 
T \int_{\mathbb{R}_+\times\mathbb{R}^d}\!\D u\D\vec{R}\: 
\left\langle \phi(t,\vec{r}+\vec{y})\phi(s,\vec{y})\wit{\phi}(u,\vec{R})^2
\right\rangle_0 
}
\nonumber \\
& & +\frac{1}{2} \int_{\mathbb{R}^{2d}}\!\D\vec{R}\D\vec{R}'\: 
a(\vec{R}-\vec{R}')
\left\langle \phi(t,\vec{r}+\vec{y})\phi(s,\vec{y})
\wit{\phi}(0,\vec{R})\wit{\phi}(0,\vec{R}') \right\rangle_0.
\label{gl:5:Crauschlos} 
\EEA
Only terms which depend
explicitly on the `noise' remain -- recall the vanishing of 
the `noiseless' two-point function 
$\langle \phi(t)\phi(s)\rangle_0=0$ because of the Bargman superselection rule. 
Using mathematical methods explained in \cite{Henk03} and restricting to
a disordered initial state $a(\vec{r})=a_0\delta(\vec{r})$, the resulting
three-point function can be calculated. It is satisfying that
$\lambda_C=\lambda_R$ follows \cite{Pico04}, in agreement with an earlier
derivation \cite{Bray94}. 

Therefore, the dynamical symmetries of non-equilibrium
kinetics are characterized by the `deterministic' part of Langevin equation.
This result only depends on the Galilei-invariance of the 
`deterministic' part (provided that $\phi$ and $\wit{\phi}$ transform
projectively through eq.~(\ref{gl:5:2:Schrpsi})). Clearly, linear
equations are Galilei-invariant in this sense. On the other hand, the question
is considerably more complicated for non-linear equations. For the purposes
of this review it is sufficient to state that deterministic non-linear 
diffusion/Schr\"odinger equations with 
$\mathfrak{age}_1$ or $\mathfrak{sch}_1$ as a dynamical symmetry 
and sufficiently general to be applicable to phase-ordering kinetics 
have been explicitly constructed \cite{Stoi05,Baum05b}. In particular, 
it has been shown that independently of the form of the potential the same
representation of $\mathfrak{age}_d$ or $\mathfrak{sch}_d$ applies, in agreement
with universality. We refer to the literature for the details. 

Since all quantities of interest will reduce to some kind of 
response function, one may calculate them from the {\it requirement that 
they transform covariantly
under the action ageing subgroup} (with Lie algebra $\mathfrak{age}_d$) 
obtained from the Schr\"odinger group when leaving out
time-translations. We shall concentrate here on the
two-time autoresponse function $R(t,s)=\langle \phi(t)\wit{\phi}(s)\rangle$ 
built from so-called quasiprimary \cite{Bela84} 
scaling operators $\phi$ and $\wit{\phi}$
which transform according to the generators of the ageing algebra. 
Then the requirement of covariance of $R$ reduces to the two conditions 
$X_0 R(t,s) = X_1 R(t,s)=0$. Since time-translations are not
included in the ageing group, the generators $X_n$ can be
generalized from (\ref{gl:5:2:SchrGen}) to the following form
\BEQ \label{gl:Xnext}
X_n = -t^{n+1}\partial_t - \frac{n+1}{2} t^n r\partial_r
-\frac{(n+1)n}{4}{\cal M}t^{n-1} r^2 - \frac{x}{2}(n+1)t^n -\xi n t^n
\;\; ; \;\; n\geq 0
\EEQ
where $\xi$ is a new quantum number associated with the field $\phi$ on 
which the generators $X_n$ act. Consequently, a quasiprimary scaling operator
which transforms covariantly under the ageing algebra is characterized
by the triplet ($x,\xi,{\cal M}$). The last term in (\ref{gl:Xnext}) can only 
be present for systems out of an equilibrium state. In particular, the 
requirement of time-translation invariance
and $[X_1,X_{-1}]=2X_0$ lead to $\xi=0$ and any representation
of $\mathfrak{sch}_d$ must have $\xi=0$. In addition, the last commutator in 
(\ref{gl:LNP:Scomm}) must be replaced by
$[{\cal S},X_1]=-2t{\cal S} -(2x+2\xi-1)M_0$. 
The meaning of $\xi$ can be understood
by integrating the generators $X_n$ in order to recover the finite,
non-infinitesimal transformations. Then it follows that if $\xi\ne 0$, lattice
observables are related to the quasiprimary fields $\Phi$ of ageing invariance
according to $\phi_{\rm lattice}(t)\mapsto \mathfrak{a}^{-x} t^{\xi} \Phi(t)$
with an additional time-dependent factor \cite{Henk06a} and where $\mathfrak{a}$
is a lattice constant. Rather than being
an exotic exception, this extension of the Schr\"odinger transformations 
seems to occur quite generically, the best-known
examples being the one-dimensional Glauber-Ising model at zero temperature. For further
examples, see table~\ref{Tabelle2} below. Finally, solving the two differential
equations for $R$ gives the explicit form of $R(t,s)$, see (\ref{gl:Rf}) below. 

While this discussion was carried out explicitly for the case $z=2$, it is
tempting to try and generalize this idea to more general values of $z$. 
In this way, the notion of {\em local scale-transformation} has been
introduced, which is based on the following main assumptions \cite{Henk02}. 
\begin{enumerate}
\item In principle, the following conformal time-transformations should be
included 
\BEQ \label{gl:5:3:Moeb}
t\mapsto t' = \frac{\alpha t + \beta}{\gamma t +\delta} \;\; ; \;\;
\alpha\delta - \beta\gamma =1
\EEQ
For applications to ageing, however, time-translations generated by $\beta$
must be left out (generalizing the restriction 
$\mathfrak{sch}_d\to\mathfrak{age}_d$). 
\item The generator $X_0$ of scale-transformations is
\BEQ \label{gl:5:3:X0gen}
X_0 = - t \partial_t - \frac{1}{z} r \partial_r - \frac{x}{z}
\EEQ
where $x$ is the scaling dimension of the quasi-primary operator on which
$X_0$ is supposed to act. Physically, this implies that there is a single 
relevant length scale $L(t)\sim t^{1/z}$. 
\item Spatial translation-invariance is required. 
\end{enumerate}
Generators for infinitesimal local scale-transformations have been explicitly
constructed \cite{Henk02}. In almost all cases, however, it turned out that
terms containing certain fractional derivatives are needed. Skipping over the
details of their construction which can be found in the literature, 
it is sufficient here to state that one can show
that the linear equation ${\cal S}_z \phi=0$ with the generalized 
Schr\"odinger operator
\BEQ
{\cal S}_z = -\lambda \partial_t + \left(\nabla_{\vec{r}}^2\right)^{z/2}
\EEQ
where $\lambda$ is a constant, 
satisfies with the generators of local scale-transformations commutation
relations quite similar to eq.~(\ref{gl:LNP:Scomm}) which establishes 
local scale-invariance as a dynamical symmetry of that linear equation 
\cite{Henk02,Roet06,Baum06e}. Attempts to suppress the fractional terms
in the generators of local scale-transformations 
lead either to $\lambda\to 0$ or to $\lambda\to\infty$ (there is an exception
if $z=1$ \cite{Henk02}). Explicit tests have recently been performed
for $z=4$ in models of kinetic growth described by the Mullins-Herring
equation \cite{Roet06} and the kinetic spherical model with conserved 
order-parameter \cite{Baum06e}. Further tests in long-ranged models
where $z$ can be continuously tuned through a control parameter are currently
being carried out \cite{Baum07a}. On the other hand, a systematic derivation
of an analogue of the Bargman superselection rules for $z\ne 2$ is still
lacking and indeed constitutes one of the most important open questions to be 
overcome for a further development of the theory. 

{\em Local scale-invariance} (LSI) assumes in particular that the two-time 
response functions transform covariantly under these local 
scale-transformations, hence $X_0R=X_1R=0$.
This leads to the following
prediction for the autoresponse \cite{Henk02,Pico04,Henk05a,Henk06a}
\BEQ \label{gl:Rf}
R(t,s) = \left\langle \phi(t)\wit{\phi}(s)\right\rangle = s^{-1-a} f_R(t,s) 
\;\; , \;\; 
f_R(y) = f_0\, y^{1+a'-\lambda_R/z} (y-1)^{-1-a'}
\EEQ
where the exponents $a,a',\lambda_R/z$ are related to $x,\xi,\wit{x},\wit{\xi}$ 
and $f_0$ is a normalization constant.\footnote{We point out that the
prediction (\ref{gl:Rf}) as well as the explicit form  (\ref{gl:Xnext}) of
$X_n$, valid for $z=2$, 
assume that the mean order-parameter $\langle \phi(0,\vec{r})\rangle=m_0=0$
at the initial moment when the quench to $T<T_c$ or $T=T_c$ is made.}
Spatio-temporal responses can be found similarly, with the result
\BEQ \label{gl:rz}
R(t,s;\vec{r}) = \left. \frac{\delta \langle
\phi(t,\vec{r})\rangle}{\delta h(s,\vec{0})}\right|_{h=0} = R(t,s) 
\Phi\left( |\vec{r}| (t-s)^{-1/z}\right)
\EEQ
where $\Phi(u)=\exp(-\frac{1}{2}{\cal M} u^2)$ if $z=2$. For 
$z\ne 2$, the function $\Phi(u)$ has to be found from a known fractional 
differential equation \cite{Henk02,Roet06,Baum06e}.

\begin{table}[t]
\caption[Tabelle1]{Magnetic systems quenched into the coexistence phase 
($T<T_c$) which satisfy (\ref{gl:Rf}) with the exponents
$a=a'$ and $\lambda_R$. $d$ is the spatial dimension and the 
numbers in brackets estimate the numerical uncertainty in the last digit(s). 
In the spherical model, long-range initial conditions are included and
in the long-range spherical model the exchange couplings decay as
$J_{\vec{r}}\sim |\vec{r}|^{-d-\sigma}$. In the bond-disordered Ising model,
the couplings are taken homogeneously from the interval $[1-\eps/2,1+\eps/2]$.
Then $z=z(T,\eps)=2+\eps/T$ \cite{Paul04} and one observes 
roughly $1.3\lesssim \lambda_R(T,\eps) \lesssim 1.7$.
\label{Tabelle1}
}
\begin{center}
\begin{tabular}{||l|rc|rr|c|l||} \hline\hline\hline
model & $d$ & $z$ & $a=a'$ & $\lambda_R$ & & Ref. \\ \hline\hline
Ising & 2 & 2 & $1/2$      & 1.26(1) & & \cite{Henk03b} \\
      & 2 & 2 & $\simeq 0.5$ & 1.24(2) & & \cite{Lore06,Jank06} \\
      & 3 & 2 & $1/2$      & 1.60(2) & & \cite{Henk03b} \\ \hline
Potts-3 & 2 & 2 & 0.49     & 1.19(3) & & \cite{Lore06,Jank06} \\\hline
Potts-8 & 2 & 2 & 0.51     & 1.25(1) & & \cite{Lore06,Jank06} \\\hline
XY      & 3 & 2 & 0.5      & 1.7     & & \cite{Abri04b} \\
XY spin wave & $\geq 2$ & 2 & $d/2-1$ & $d$ & angular response & \cite{Pico04} \\\hline
spherical & $>2$ & 2 & $d/2-1$ & $(d-\alpha)/2$ & $C_{\rm ini}(\vec{r})\sim |\vec{r}|^{-d-\alpha}$ & \cite{Newm90,Pico02} \\\hline
long-range  & $>2$ & $\sigma$ & $d/\sigma-1$ & $d/2$ & $0<\sigma<2$ & \\
spherical & $\leq 2$ & $\sigma$ & $d/\sigma-1$ & $d/2$ & $0<\sigma<d$ & \cite{Cann01} \\\hline\hline
random-bond Ising & 2 & $2+\eps/T$ & $1/z(T,\eps)$ & $\lambda_R(T,\eps)$ & disordered & \cite{Henk06b}
\\\hline\hline\hline
\end{tabular}\end{center}
\end{table}

\begin{table}
\caption[Tabelle 2]{Systems quenched to a critical point of their
stationary state which satisfy (\ref{gl:Rf}) with the exponents $a$, $a'$ 
and $\lambda_R/z$. $d$ is the spatial dimension and the 
numbers in brackets estimate the uncertainty in the last digit(s). 
{\sc csm} stands for the spherical model with a conserved order-parameter,
{\sc fa} denotes the Frederikson-Andersen model, {\sc nekim} is the
non-equilibrium kinetic Ising model and {\sc bcp} and {\sc bpcp} denote the
bosonic contact and pair-contact processes, respectively. In the {\sc bpcp} 
dynamical scaling only holds along a part of the critical line.  
In the spherical model, long-range initial correlations 
$C_{\rm ini}(\vec{r})\sim |\vec{r}|^{-d-\alpha}$ were considered. 
If $d+\alpha>2$, these reduce to short-ranged initial correlations 
(denoted {\sc s}), but for $d+\alpha<2$ a new class {\sc l} arises. 
In those models described by a Langevin equation, the simple white
noise $\langle\eta(t,\vec{r})\eta(s,\vec{r}')\rangle
=2T\delta(\vec{r}-\vec{r}')\delta(t-s)$ was used, with 
the only exception of the {\sc csm}. 
In the Ising spin glass, a bimodal disorder was used. 
\label{Tabelle2}
}
\begin{center}
\begin{tabular}{||l|r|rrr|cl||} \hline\hline\hline
model & \multicolumn{1}{c|}{$d$} & $a$ & $a'-a$  & $\lambda_R/z$ & & Ref.  \\\hline\hline
random walk & & -1 & 0 & 0 & & \cite{Cugl94b} \\ \hline
OJK-model & & $(d-1)/2$ & $-1/2$  & $d/4$ &  & \cite{Bert99,Maze04,Henk05a}\\\hline
Ising & 1 & 0 & $-1/2$  & $1/2$ &  & \cite{Godr00a,Lipp00,Henk03d}\\ 
      & 2 & $0.115$ & $-0.187(20)$ & $0.732(5)$ & & \cite{Plei05,Henk06a}\\ 
      & 3 & $0.506$ & $-0.022(5)$  & $1.36$     & &\cite{Plei05,Henk06a}\\\hline
XY    & 3 & 0.52    & 0            & 1.34(5)    & & \cite{Abri04b} \\\hline
spherical $d>2$ & $<4$ & $d/2-1$ & 0 & $d/2-\alpha/4-1/2$ & {\sc l} & \cite{Pico02} \\
          & $>4$ & $d/2-1$ & 0 & $(d-\alpha)/4 +1/2$ & {\sc l} &
            \cite{Pico02}\\
          & $<4$ & $d/2-1$ & 0 & $3d/4-1$ & {\sc s} & \cite{Godr00b} \\
	  & $>4$ & $d/2-1$ & 0 & $d/2$ & {\sc s} & \cite{Godr00b} \\\hline\hline
{\sc csm} & $>2$ & $d/4-1$ & 0 & $(d+2)/4$ &   & \cite{Baum06e} \\ \hline\hline
disordered Ising     & $4-\eps$ & $1-\frac{1}{2}\sqrt{\frac{6\eps}{53}}$ & 0 & $3-\frac{1}{2}\sqrt{\frac{6\eps}{53}}$ & O($\eps$), $\log$ & \cite{Cala02a,Sche05,Sche06}\\\hline\hline
{\sc fa} & $>2$ & $1+d/2$ & $-2$  & $2+d/2$ &  & \cite{Maye06} \\
         & $1$ & $1$ & $-3/2$ & $2$ &  & \cite{Maye06,Maye04}\\\hline
Ising spin glass & 3 & $0.060(4)$ & $-0.76(3)$ & $0.38(2)$ & see sect. 4 & \cite{Henk05a,Henk05b} \\ \hline\hline 
contact process  & 1 & $-0.681$ & $+0.270(10)$ & $1.76(5)$ & $t/s\gtrsim 1.1$ & \cite{Enss04,Hinr06,Henk06a} \\
 & $>4$ & $d/2-1$ & 0 & $d/2+2$ & & \cite{Rama04} \\\hline
{\sc nekim} & 1        & -0.430(4) & 0 & 1.9(1)& & \cite{Odor06} \\\hline
{\sc bcp}   & $\geq 1$ & $d/2-1$ & 0   & $d/2$ & & \cite{Baum05a,Baum05b} \\ \hline
{\sc bpcp}  & $>2$     & $d/2-1$ & 0   & $d/2$ & $\alpha\leq\alpha_C$ & \cite{Baum05a,Baum05b} \\\hline\hline\hline 
\end{tabular}\end{center}
\end{table}

Starting with \cite{Henk01}, the prediction (\ref{gl:Rf}) has been reproduced 
in many different spin systems
and we list examples quenched to below criticality in table~\ref{Tabelle1}
and quenched to the critical point in table~\ref{Tabelle2}. For $T<T_c$,
it is found empirically that $a=a'$ in all examples considered so far.\footnote{Recall that $M_{\rm TRM}(t,s)$ in direct space
is not very sensitive to $a-a'$ and that most numerical studies were carried
out in this setting. A considerably more sensitive test looks at 
$M_{\rm TRM}$ in momentum space \cite{Plei05}.} We point out that agreement 
with local scale-invariance eq.~(\ref{gl:Rf}) is not
only obtained for systems where the dynamical exponent is $z=2$, but that
rather there exist quite a few examples where $z$ can become 
considerably larger or smaller than 2. It must be remembered, 
however, that the above 
derivation of (\ref{gl:Rf}) for a stochastic Langevin equation has for the time 
being only been carried out for $z=2$ and the justification of
$X_0R=X_1R=0$ remains an open problem for $z\ne 2$ in general although the 
result eq.~(\ref{gl:Rf}) seems to work remarkably well.  
It is non-trivial that a relatively simple
extension of dynamical scaling should be capable of making predictions
which can be reproduced in physically quite different systems. 

A few comments are still needed: (i) for the XY model in the 
spin-wave approximation (table~\ref{Tabelle1}), 
eq.~(\ref{gl:Rf}) holds for the response of the 
angular variable $\phi=\phi(t,\vec{r})$ which is related to the XY spin through
$\vec{S}=(\cos \phi,\sin \phi)$. Magnetic responses have a different scaling
form \cite{Bert01,Abri04a}. (ii) In the critical disordered Ising model
(table~\ref{Tabelle2}) one finds a logarithmic scaling form
$R(t,s)=(r_0+r_1\ln (t-s))f_R(t/s)$ \cite{Cala02a,Sche05,Sche06} such that
the computed $f_R(y)$ is consistent with (\ref{gl:Rf}) to one-loop
order, or up to terms of order ${\rm O}(\eps)$. (iii) Finally, a two-loop
calculation of the critical non-conserved O($n$)-model does produce in 
$4-\eps$ dimensions an expression for $f_R(y)$ which is incompatible
with (\ref{gl:Rf}) \cite{Cala02} and a similar result is anticipated in
$2+\eps$ dimensions \cite{Fedo06}, although the one-loop results are still
compatible \cite{Cala01,Cala02,Fedo06}. Should one conclude from these studies 
that for $T=T_c$ the prediction (\ref{gl:Rf}) and by implication local 
scale-invariance can only hold approximatively~? 
This might well be a subtle question. Deviations
between (\ref{gl:Rf}) and the field-theoretical studies typically arise when 
$t/s\approx 1$. However, in this region the field-theoretical 
results for $f_R(y)$ do not
agree with the ones of non-perturbative numerical studies \cite{Plei05}. 
Since the perturbative expansion usually carried out in field-theoretical
studies does not necessarily take care of the Galilei-invariance, it is
necessary to carefully check that the truncation of the $\eps$-series does
not introduce slight inaccuracies. Only after this has been done (for example
by re-summing the $\eps$-series) and checked by comparing with
non-perturbative data, meaningful quantitative statements on the 
scaling functions can be made. (iv) Throughout, it was implicitly assumed that
the order-parameter vanishes initially. Systematic studies on what
happens when this condition is relaxed are only now becoming available
\cite{Anni06,Cala06,Fedo06,Baum06d}. 
These extensions might be particularly important 
for chemical kinetics \cite{Henk07a} and the existing simulations for directed 
percolation (or the contact process) may well turn out to be an example 
where the initial non-vanishing value of the order-parameter influences 
the form of scaling functions such as $f_R(y)$ 
\cite{Baum06d,Enss04,Rama04,Hinr06}. 
{}From that point of view it is surprising that eq.~(\ref{gl:Rf}) could 
describe any part of the data of $f_R(y)$ as well as it does. 
(v) We did not include growth models since their analysis from the point of 
view of LSI is just beginning \cite{Roet06,Baum06e}.

If $z=2$, it is also possible, using eq.~(\ref{gl:5:Crauschlos}), to derive
explicit predictions for the two-time correlation 
function \cite{Pico04,Henk04b}. These have been tested in some 
exactly solvable models \cite{Pico04,Henk06a}, the two-dimensional Ising model 
\cite{Henk04b} and the
two-dimensional $q$-states Potts model with $q=2,3,8$ \cite{Lore06,Jank06}. 
Extensions to $z=4$ have been studied very recently \cite{Roet06,Baum06e}. 

Finally, let us mention that the prediction (\ref{gl:rz}) for the space-time 
response has been verified for $z=2$ through numerical simulations of two- and 
three-dimensional Ising models undergoing phase-ordering \cite{Henk03e} and 
{}from the exact solutions of the $1D$ Glauber-Ising model 
\cite{Godr00a,Lipp00,Henk03d} at $T=0$, the
spherical model in $d>2$ dimensions \cite{Godr00b,Pico02} and the 
long-range spherical model \cite{Cann01,Baum07a}.

\section{Disordered ferromagnets}

We now take up the discussion of disordered systems. Rather than going
directly to spin glasses, it may be useful to consider first disordered, 
but not frustrated systems, such that the random exchange couplings 
$J_{i,j}\geq 0$ in models such as (\ref{gl:disI}). 

We point out that a disorder in the exchange couplings, as introduced
in the random-bond model (\ref{gl:disI}), can have quite different 
consequences than a
random site dilution, as described by a classical hamiltonian
${\cal H} = - J \sum_{(i,j)} \varepsilon_i\varepsilon_j \sigma_i \sigma_j$
where $\varepsilon_i\in\{0,1\}$ are random variables selected according 
to a control parameter $p$. Provided that on average
$\overline{J_{i,j}}=1$, the critical temperature in the presence of bond
disorder will not change much. For a site-diluted model, however, the
phase-transition should disappear once the non-magnetic sites (where
$\varepsilon_i=0$) will begin to percolate. Here, we shall mainly restrict
attention to random-bond disorder and shall furthermore restrict to the 
case of quenched, that is immobile, disorder.

The first question to be addressed concerns the existence and the detailed
form of dynamical scaling, as expressed through the time-dependence $L=L(t)$
of the linear size of the clusters. The physical picture is as follows
\cite{Huse85}, see also the middle column of figure~\ref{LNPHePl:amas3fois}. 
Initially, small clusters will form and start to grow. These
early stages are still unaffected by the presence of the disorder which will 
only begin to be felt once the average cluster size has become of the same
order as the mean distance between two impurities. Then the domain walls
will become trapped close to the disorder-induced defects and their motion 
will slow down correspondingly. Rather than being described by purely
geometric considerations as it can be done through the Allen-Cahn equation
for non-disordered systems, the impurities act as energy barriers to the 
domain growth, hence the pinning centers are localized in energetically
favourable positions, which explains the slowing-down of the kinetics. 
Phenomenologically, for a non-conserved order-parameter one still expects
that the basic curvature-reducing mechanism is applicable, leading to 
\cite{Huse85,Lai88}. This yields
\BEQ \label{gl:wachs}
\frac{\D L(t)}{\D t} = \frac{D(L,T)}{L(t)}
\EEQ
where the diffusion constant $D=D(L,T)$ now depends on the domain scale 
$L=L(t)$
and the temperature $T$. For a constant $D$, one is back to the 
non-disordered case, with $L(t)\sim t^{1/2}$. For thermally activated motion, 
one should have
\BEQ
D(L,T) \simeq D_0 \exp(-E_B/T)
\EEQ
where $E_B=E_B(L)$ is the barrier energy. Huse and Henly \cite{Huse85}
have argued long ago that the barrier energy should depend on the domain size 
algebraically $E_B(L)\simeq E_{0} L^{\psi}$,
where the exponent $\psi=(2\zeta+d-3)/(2-\zeta)$ where $\zeta$ is the
domain-wall roughness exponent. For example, in two dimensions it is known that
$\zeta=\frac{2}{3}$, hence $\psi=\frac{1}{4}$. Inserting into
eq.~(\ref{gl:wachs}) leads to \cite{Huse85}
\BEQ \label{hh}
L(t) = \left(\frac{T}{E_0}\right)^{1/\psi} {\cal L}\left(\frac{t}{t_0}\right)
\;\; , \;\; 
{\cal L}(\tau) = \left\{ 
\begin{array}{ll} \frac{2}{\psi} \tau & \mbox{\rm ~~;~ $\tau\ll 1$} \\
(\ln \tau)^{1/\psi} & \mbox{\rm ~~;~ $\tau\gg 1$}
\end{array} \right.
\EEQ
which describes the qualitative change of behaviour between the initial
regime and a second regime of slower growth for larger times.    

\begin{table}
\caption[Exp growth]{Experimental results on the growth of the domain size
$L(t)$ in some disordered, but non-glassy systems in two dimensions. 
For Rb$_2$Co$_{0.60}$Mg$_{0.40}$F$_4$ a different temperature-dependent 
exponent $\psi$ is found close to $T_c$.  
For {\sc tgs} the order-parameter is conserved.\label{Tabelle3}}
\begin{tabular}{|l|l|ll|r|l|} \hline\hline
material & model & $L(t)$ & $\psi$ & $T_c \mbox{\rm [$K$]}$ & Ref. \\ \hline\hline
Rb$_2$Co$_{0.60}$Mg$_{0.40}$F$_4$ & diluted Ising & $A+B (\ln t)^{1/\psi}$   & $0.28 (15\mbox{\rm [$K$]}/T)$ &  20  & \cite{Iked90} \\ 
                                & antiferromagnet & &  & & \\ 
				& & \multicolumn{2}{r|}{if $T\lesssim 
				  15\mbox{\rm [$K$]}$~} & & \\ \hline
tryglycine sulfate              & ferroelectric & \multicolumn{2}{l|}{$\sim (t-t_0)^{1/z}$ ~~~ $z\approx 3$} & 322 & \cite{Liko00,Liko01} \\
({\sc tgs})                     & {\em conserved} & $\sim (\ln t)^{1/\psi}$ & $1/4$   &     & \\ \hline
Rb$_2$Cu$_{0.89}$Co$_{0.11}$F$_4$ & random-bond & $(\ln t)^{1/\psi}$ & $0.20(5)$ & 4.93(5) & \cite{Schi93} \\ 
                                & Ising & \multicolumn{2}{r|}{$T=3.9 \mbox{\rm [$K$]}$~}       & & \\ \hline
    
ZLI 4792 (Merck)                & random-bond & $(\ln t)^{1/\psi}$ & $1/4$ & & \cite{Shen99} \\ 
(liquid crystal)                & Ising  & & & &\\ \hline\hline 
\end{tabular}
\end{table}

We now compare this result with the available experimental evidence collected
in table~\ref{Tabelle3}. Although there is a general 
qualitative agreement about a a change from a relatively fast kinetics seen 
at not too late times \cite{Iked90,Liko00,Liko01} to slowing-down of the 
kinetics at later times, quantitatively the situation remains a little 
ambiguous. 
In the published work known to us, data were usually compared to a 
logarithmic law $L(t)\sim (\ln t)^{1/\psi}$, which rather than being an 
objective statement might simply come from the fact that the prediction 
(\ref{hh}) of Huse and Henley \cite{Huse85} has been around for quite 
some time. On the
other hand, there is some experimental information which is not consistent
with eq.~(\ref{hh}), if it can be taken at face value. 
For example, in at least one system the effective exponent depends 
on temperature \cite{Iked90} while
a constant $\psi=\frac{1}{4}$ is theoretically expected in two dimensions 
\cite{Huse85}. Indeed, it has been
argued that in site-diluted Ising models, because of the fractal nature of the
domains boundaries the energy barriers should rather depend logarithmically 
on $L(t)$, viz. $E_B(L)=\eps \ln(1+L)$ \cite{Henl85,Ramm85,Paul04}. 
Inserting this into (\ref{gl:wachs}) leads to an algebraic law 
$L(t)\sim t^{1/z}$ where the effective $z$
crosses over from $z=\frac{1}{2}$ for short times to 
\BEQ \label{eq:z}
z = z(T,\eps) = 2 + \eps/T
\EEQ
for late times. In at least one experiment, data were explicitly seen to be 
also compatible with an algebraic growth law \cite{Liko00,Liko01}. 

For a conserved order-parameter, a similar discussion can be carried out. 
It can be shown that for algebraic energy barriers the result (\ref{hh}) 
remains unchanged \cite{Huse85}, whereas for logarithmic barriers one finds 
again algebraic growth $L(t)\sim t^{1/z}$ with $z=z(T,\eps)=3+\eps/T$ 
\cite{Paul04,Paul05}. 

Theoretical work has among others studied a random-bond Ising model defined by 
the classical hamiltonian
\BEQ \label{gl:rbI}
{\cal H}_{\rm dis} = - \sum_{(i,j)} J_{i,j} \sigma_i \sigma_j
\EEQ
where the $J_{i,j}$ are random variables uniformly distributed over the
interval $[1-\varepsilon/2, 1+\varepsilon/2]$ and where a non-conserved 
dynamics was created via a Metropolis or heat-bath algorithm. For
examples, estimates for $L(t)$ extracted from the single-time 
spatio-temporal correlation function were seen to be in full agreement 
with the prediction (\ref{eq:z}) of logarithmic energy barriers 
\cite{Paul04,Paul05}. 
The same conclusions are obtained from the dynamical scaling behaviour of 
two-time quantities \cite{Henk06b}, see also below. 
It would be desirable to compare existing
experimental data with the possibility of logarithmic energy barriers. If the
discrepancy between the information extracted from experiments and theoretical
simulations should persist, it could mean that the usually considered
theoretical models are not realistic enough for a quantitative description of
real disordered materials.

We here discuss the thermoremanent magnetization $M_{TRM}(t,s)$ which 
is obtained 
when quenching the system in presence of a small magnetic field. 
The magnetic field is cut after the waiting time $s$ and the decay of the 
magnetization is then monitored
as a function of time. The thermoremanent magnetization is related 
to the response $R(t,s)$ through
\BEQ 
\label{gl:Mtrm}
M_{\rm TRM}(t,s) = h_0 \int_{0}^{s} \!\D u\: R(t,u),
\EEQ
where $h_0$ is the amplitude of the small magnetic field.
The interest in this integrated response function comes from the fact that 
a direct calculation
of the functional derivative which defines $R$ produces extremely noisy data, 
in contrast to the
measurement of $M_{TRM}$ where the noise is to a large extend smoothed out 
by the integration.

{}From our experience with the scaling  of $M_{\rm TRM}(t,s)$ in the 
phase-ordering of simple magnets, we expect the
scaling behaviour \cite{Henk02a,Henk03e}
\BEQ \label{gl:M}
M_{\rm TRM}(t,s) = r_0 s^{-a} f_M(t/s) + r_1 s^{-\lambda_R/z} g_M(t/s) 
\;\; , \;\;
g_M(y) \simeq y^{-\lambda_R/z}.
\EEQ
Note that we have here included the leading correction term 
$\sim s^{-\lambda_R/z} g_M(t/s)$
which can become quite sizeable in systems undergoing phase-ordering 
and which must be
subtracted off before a reliable determination of the scaling 
function $f_{M}(y)$ is possible.
In the framework of LSI an analytical expression for $f_{M}(y)$ is readily 
derived from (\ref{gl:Rf})
with $a=a'$ and reads (${}_2F_1$ is a hypergeometric function)
\BEQ \label{gl:LSI}
f_{M}(y) = y^{-\lambda_R/z} {}_2F_1\left( 1+a, \frac{\lambda_R}{z}-a;
\frac{\lambda_R}{z}-a+1; \frac{1}{y} \right).
\EEQ

\begin{figure}[!h] 
\centerline{\psfig{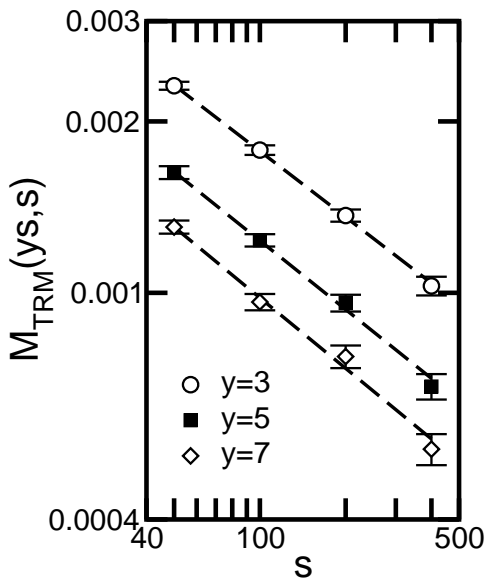}}
\caption{Thermoremanent magnetization $M_{\rm TRM}(ys, s)$ of the two-dimensional
random-bond Ising model (\ref{gl:rbI}) as a function
of $s$ for $\eps=0.5$ and $T=1.0$ and different values of $y$.
\label{fig:trm_disorder} }
\end{figure}

The expected scaling behaviour (\ref{gl:M}) of the linear response has been 
studied through 
large-scale numerical simulations \cite{Henk06b}. As an example, figure 
\ref{fig:trm_disorder} shows data obtained
for $\varepsilon=0.5$ and $T=1$. This is a case where the finite-time 
correction to the leading
behaviour $\sim s^{-a} f_M(t/s)$ is negligible, i.e. $r_1 \approx 0$ in 
eq.~(\ref{gl:M}).
Plotting $M_{\rm TRM}(ys, s)$ over against $s$ in a log-log plot nicely 
verifies 
the expected scaling behaviour and yields in addition an estimate for the 
ageing exponent $a$,
see table~\ref{tabl4}. It is worth noting that the estimates are in complete 
agreement with the
relation $a(T,\epsilon)=1/z(T,\epsilon)$ expected for systems of class $S$ as 
discussed in section~1,
where the dynamical exponent $z$ is given by (\ref{eq:z}). This confirms the 
conclusion of 
\cite{Paul04,Paul05} that the random-bond Ising model shows simple ageing and 
furthermore $L(t)\sim t^{1/z}$ with $z=z(T,\epsilon)$
given by  eq.~(\ref{eq:z}). 

We also compiled in table~\ref{tabl4} the estimates \cite{Henk06b} 
for the non-universal constants $r_0$ and $r_1$
as well as for the exponent $\lambda_R/z$ which governs the asymptotic 
behaviour of the
scaling function $f_M(y) \sim y^{-\lambda_R/z}$.

\begin{table}[!h] 
\caption[Exponenten]{Quantities describing the dynamical scaling behaviour of
the linear response $M_{TRM}$ in the bond-disordered two-dimensional
Ising model for different values of $\epsilon$ and different temperatures $T$.
The last column is obtained from equation (\ref{eq:z}) and should be compared 
with the numerical estimates for the exponent $a$.
\label{tabl4}}
\begin{center}
\begin{tabular}{|cc|ll|cc|c|} \hline\hline
$\eps$ & $T$ &  \multicolumn{1}{|c}{$a$} & $\lambda_R/z$ & $r_0$    & $r_1$     & $1/z$ \\ \hline\hline
0.5    & 1.0 & 0.398(5)   & 0.61(1)       & 0.021(1) & 0         & 0.400 \\
       & 0.8 & 0.382(4)   & 0.595(10)     & 0.020(1) & 0         & 0.381 \\
       & 0.6 & 0.353(4)   & 0.58(1)       & 0.022(1) & 0         & 0.353 \\
       & 0.4 & 0.310(5)   & 0.52(1)       & 0.029(2) & 0.008(1)  & 0.308 \\ \hline
1.0    & 1.0 & 0.330(5)   & 0.51(1)       & 0.021(1) & -0.009(1) & 0.333 \\
       & 0.8 & 0.308(4)   & 0.49(1)       & 0.019(1) & -0.006(1) & 0.308 \\
       & 0.6 & 0.277(6)   & 0.46(1)       & 0.020(1) & -0.007(1) & 0.273 \\
       & 0.4 & 0.22(1)    & 0.375(10)     & 0.026(2) & -0.014(3) & 0.222 \\ \hline
2.0    & 1.0 & 0.24(2)    & 0.33(1)       & 0.048(2) & -0.048(4) & 0.250 \\
       & 0.8 & 0.22(2)    & 0.30(1)       & 0.093(3) & -0.042(4) & 0.222 \\
       & 0.6 & 0.17(2)    & 0.27(1)       & 0.194(4) & -0.033(3) & 0.188 \\ \hline\hline
\end{tabular}\end{center}
\end{table}

\begin{figure}[!h] 
\centerline{\psfig{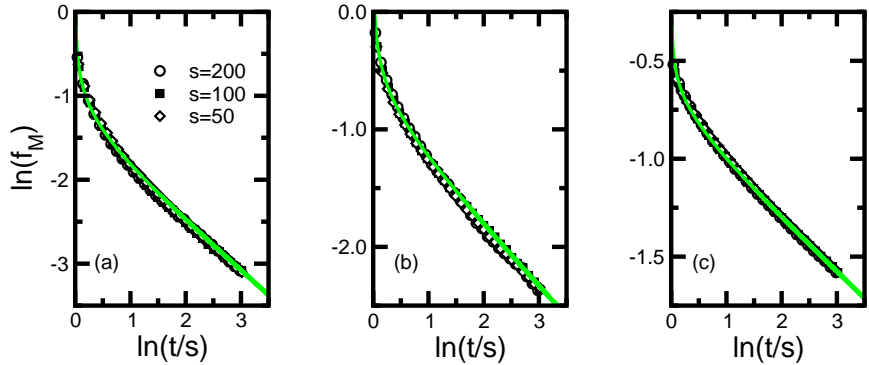}}
\caption{Comparison of the numerically determined scaling functions $f_M(t/s)$
of the two-dimensional random-bond Ising model with the LSI prediction 
eq. (\ref{gl:LSI}) (full lines) for (a) $\varepsilon=0.5$ and $T=0.6$,
(b) $\varepsilon=1$ and $T=1$ and
(c) $\varepsilon=2$ and $T=0.6$.
\label{fig:LSI_disorder}}
\end{figure}

The random-bond Ising model is very well suited for investigating in detail 
the possible applicability of 
the theory of local scale-invariance to nontrivial systems undergoing 
phase-ordering.
This is of course due to the fact that the dynamical exponent can be 
continuously changed
by changing the distribution of the couplings and/or the 
temperature, yielding values
of $z$ much larger than the value $z=2$ encountered in 
non-disordered ferromagnets
undergoing phase-ordering. We have seen in section~2 for what models 
the data for $M_{\rm TRM}(t,s)$ agree with LSI, see table~\ref{Tabelle1}. 
Figure~\ref{fig:LSI_disorder} compares for three different cases the computed 
scaling function $f_M$ of the thermoremanent magnetization
with the LSI prediction (\ref{gl:LSI}). The same comparison was made for
$\varepsilon=0.5, T=1$, $\varepsilon=1,T=0.6$, and $\varepsilon=2,T=1$ in our original 
paper \cite{Henk06b}. It has to be recalled that prior to 
this comparison
the values of the exponents entering into (\ref{gl:LSI}) and also the 
normalization
are already fixed. We see that the form of the scaling function (after 
subtraction
of the above-mentioned finite-time correction, if needed) 
is perfectly described by LSI,
and this is the case for all values of $\varepsilon$ and $T$ investigated. 
This is a very remarkable result as it suggests that the idea of extending 
dynamical
scaling to local, space- and time-dependent, scaling is capable of reproducing 
faithfully
the linear responses of very different systems characterized by very different 
values of the dynamical exponent.

Having looked at the evidence in favour of simple scaling of $M_{\rm TRM}(t,s)$,
it may be useful here to take a slightly broader point of view. Indeed, 
in the discussion of glassy systems, people often describe the scaling
of, say, the two-time correlation function as follows, e.g. \cite{Vinc06}
\BEQ \label{eq:ag1}
C(t,s) = C_{\rm st}(t-s) + C_{\rm age}(t,s)
\EEQ
such that the stationary part satisfies $\lim_{t\to\infty} C_{\rm st}(t)=0$ 
and furthermore \\ 
$\lim_{t-s\to\infty} (\lim_{s\to\infty} C_{\rm st}(t-s))=q_{\rm EA}$,
where $q_{\rm EA}$ is the Edwards-Anderson order-parameter for glasses. On the
other hand, the ageing part is assumed to read
\BEQ \label{eq:ag2}
C_{\rm age}(t,s) = {\cal C}\left(\frac{h(t)}{h(s)}\right) \;\; , \;\;
h(t) = h_0 \exp\left[ \frac{1}{A}\frac{t^{1-\mu}-1}{1-\mu}\right]
\EEQ
where ${\cal C}$ is a scaling function, $\mu$ is a free parameter and
$h_0$ and $A$ are constants. 
Conventionally, one refers to the case $0<\mu<1$ as {\em subageing}, the
limit case $\mu\to 1$ as (full or simple) {\em ageing} and the case $\mu>1$ as
{\em superageing}. However, it has been shown by Kurchan \cite{Kurc02} 
that, given that the positive function ${\cal C}(u)$ decreases 
strictly monotonously
with $u$, the case $\mu>1$ is incompatible with an elementary property of the
autocorrelation function, namely that if there is a strong correlation
between times $t_1$ and $t_2>t_1$ and as well a strong correlation between times
$t_2$ and $t_3>t_2$, then there must exist a strong correlation between times
$t_1$ and $t_3$. Hence {\it superageing is impossible} \cite{Kurc02}. 

There is a nice argument which explains the origin of the form
(\ref{eq:ag1},\ref{eq:ag2}) \cite{Andr06}. It relies on the
observation that if $C(t,s)$ is plotted over against $t-s$, a plateau is
observed, see also figure~\ref{LNPHePl:Abb2}a. Further, according to 
Zippold, K\"uhn et Horner \cite{Zipp00} the transition
towards the scaling regime occurs at a time difference $t-s\sim t^{\zeta}$
where $0<\zeta<1$ describes this change of behaviour.\footnote{Explicitly,
$\zeta=4/(d+2)$ in the $d$-dimensional spherical model and $\zeta=4/5$ in the
spherical spin glass \cite{Zipp00}.} Hence close to the plateau one can
assume
\BEQ \label{eq:agC1}
C(t,s) = q_{\rm EA} + t^{-\alpha} g_1\left((t-s) t^{-\zeta}\right) + \ldots 
\EEQ
where $\alpha$ is some exponent and $g_1$ a scaling function. Now, consider
$t-s=x t^{\zeta}$ and take the limit $t\to\infty$ and then $x\gg 1$. Then, 
to leading order 
\BEQ \label{eq:agC2}
C(t,s) = {\cal C}\left(\frac{h(t)}{h(s)}\right) 
\simeq {\cal C}\left( 1 + x t^{\zeta}\frac{\D \ln h(t)}{\D t}\right) 
\simeq q_{\rm EA} + c_{\rm age}^{(1)} 
\left(x t^{\zeta}\frac{\D \ln h(t)}{\D t}\right)^{\mathfrak{b}}  
\EEQ
where $\mathfrak{b}$ is a further exponent which can be worked out for 
certain model spin glasses \cite{Andr06}. Comparing
eqs.~(\ref{eq:agC1},\ref{eq:agC2}), the dependences on $t$ and on $x$ can
be separated which in particular leads to $\D \ln h(t)/\D t = A^{-1} 
t^{-\mu}$, where $\mu=\zeta+\alpha/\mathfrak{b}$ and $A$ is a separation 
constant. Then eq.~(\ref{eq:ag2}) follows directly
\cite{Andr06}. In particular, this means that other phenomenological forms
which have been discussed in the past can be eliminated.  
    
\begin{figure}[!ht] 
\centerline{\psfig{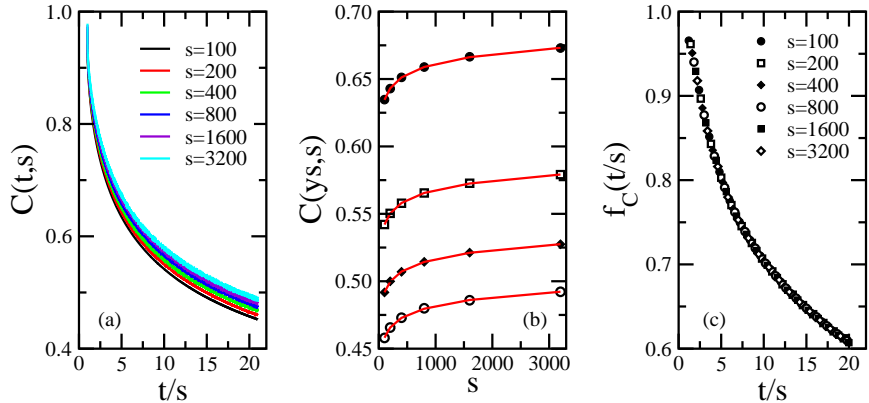}}
\caption{(a) Autocorrelation $C(t,s)$ vs $t/s$ for various waiting times $s$ 
in the random-bond Ising model with $\varepsilon=2$ and $T=1$. 
(b) Plot of $C(y s,s)$ for the same parameters and $y = 5, 10, 15, 20$
(from top to bottom). The data can be perfectly fitted by the scaling form 
(\ref{autocorr_ft}) with $b'=0.075$, which illustrates the existence 
of important finite-time corrections. 
(c) The scaling function $f_C(t/s)$ obtained after subtracting off the
finite-time corrections from the autocorrelation, again for $\varepsilon
=2$ and $T=1$, in agreement with simple ageing of the autocorrelation.
\label{fig:autocorr_disorder} }
\end{figure}

After these preparations we turn to a dicussion of ageing properties 
of the autocorrelation function in disordered ferromagnets. Indeed,
recent studies of $C(t,s)$ in
bond-diluted \cite{Henk06b} and in site-diluted \cite{Paul06} 
Ising models quenched to 
below the critical points showed surprising and at first sight unexpected features.
In figure~\ref{fig:autocorr_disorder}a we show data \cite{Henk06b} 
for the two-dimensional random-bond Ising model plotted over against $t/s$. In clear
contrast with non-disordered simple magnets, no clear scaling is seen. At first
sight, one might be tempted to introduce a non-vanishing ageing exponent $b$, 
see eq.~(\ref{gl:sC}), but since the curves for $C(t,s)$ increase with $s$, the
effective values of $b=b_{\rm eff}$ fitted to the data will turn out to be
negative \cite{Henk06b,Paul06}. If true, that would imply that
for $s\to\infty$ the autocorrelator $C(t,s)$ would grow unboundedly, which is
impossible. It is therefore necessary to reexamine whether the standard scaling 
form of simple ageing is applicable. Indeed, in a recent study of the
randomly site diluted Ising model \cite{Paul06} the more general scaling 
form (\ref{eq:ag1},\ref{eq:ag2}) was used and fitted values of $\mu$ in the
range $\mu\approx 1.03-1.04$ were reported, implying the so-called
superageing behaviour \cite{Paul06}. Although a good collapse of the data can be
obtained this way, the arguments raised by Kurchan
\cite{Kurc02} and quoted above make it doubtful that the conclusion of
\cite{Paul06} can be accepted
at face value. Indeed, in using scaling forms such as (\ref{gl:sC}) one has to 
be careful about the possibility of finite-time corrections to scaling. 
We illustrate the presence of strong corrections to scaling in 
figure \ref{fig:autocorr_disorder}b \cite{Henk07b} 
for the case $\varepsilon=2$ and $T=1$,
for several values of $y=t/s$. Clearly, the data are perfectly
described by the extended scaling form
\BEQ \label{autocorr_ft}
C(y s,s) = f_C(y) - s^{-b'} g_C(y)
\EEQ
(hence with $b=0$) and with $b' > 0$.\footnote{For some other values 
of $\varepsilon$
and of $T$ two correction terms have to be included in order to describe 
the non-monotonic variation of $C(y s,s)$ as a function of $s$, see
\cite{Henk07b} for details.} Subtracting off the leading correction term, 
yields a perfect scaling behaviour eq.~(\ref{gl:sC}) 
according to simple ageing as shown in 
figure~\ref{fig:autocorr_disorder}c \cite{Henk07b}, quite analogous 
with the standard 
scaling expected for simple magnets quenched to below their critical point.

In conclusion, the random-bond Ising model apparently ages in much the same
way as a simple magnet. The prediction of table~\ref{TabExp1} for the
exponents $a,b$ in class S systems hold true for this non-glassy disordered
system as well, but the value of the dynamical
exponent $z=z(T,\varepsilon)$ (and also of $\lambda_{C,R}$) becomes dependent
on temperature and on the distribution of the couplings. The form
of the response function agrees with the prediction of LSI, for a large
range of values of $z$. It would be 
very interesting to check these conclusions in other systems.

\section{Critical Ising spin glasses}

The disordered ferromagnet studied in the previous section can be viewed
as intermediary between the simple, non-disordered magnets and the disordered
and highly frustrated spin glasses. In the following we discuss to what
extend the results obtained for the disordered ferromagnets can be extended
to glassy systems. We thereby shall concentrate on the Ising spin glass, with a
static Hamiltonian ${\cal H} = - \sum_{(i,j)} J_{i,j} \sigma_i \sigma_j$.
Here $\sigma_i=\pm 1$ are the usual Ising spins and the nearest-neighbour
couplings $J_{i,j}$ are random variables. We shall consider three different
distributions of the couplings: (i) the {\bf bimodal distribution} with
\begin{equation}
P_{B}(J_{i,j}) = [\delta (J_{i,j}-J)+\delta (J_{i,j}+J)]/2,
\label{binomial}
\end{equation}
(ii) the {\bf Gaussian distribution} with
\begin{equation}
P_{G}(J_{i,j}) = \exp(-J_{i,j}^2/2 J^2)/(J\sqrt {2 \pi})
\label{gaussian}
\end{equation}
and (iii) the {\bf Laplacian distribution} with
\begin{equation}
P_{L}(J_{i,j}) = \exp(-\sqrt{2}\mid J_{i,j}/J \mid)/(J \sqrt{2}).
\label{laplacian}
\end{equation}
All distributions are symmetric with zero mean and variance 
$\left< J_{i,j}^2 \right>/J^2 = 1$.
The dynamics of the model is given by a master equation where the
rates are chosen according to heat-bath dynamics. 
It is by now established, see e.g. \cite{Kawa04} for a review, that for $d>2$ 
dimensions this model undergoes an
equilibrium phase-transition between a paramagnetic and a frustrated 
spin-glass phase. There has been considerable
debate on the precise relationship between the relevant time and length
scales for quenches below
the spin-glass critical temperature $T_c$. It has been attempted to summarize 
the present state of knowledge into the form \cite{Bouc01}
\BEQ \label{gl:tL}
t(L) \sim L^z \exp\left(
\frac{\Delta_0}{T} \left(\frac{L}{\xi(T)}\right)^{\psi}
\right)
\EEQ
where $\Delta_0$ is an energy scale of order $T_c$, $\psi$ is a barrier
exponent and $\xi(T)$ is the equilibrium correlation length at 
temperature $T$. This form has been
used to fit successfully simulational data in the three-dimensional and 
in the four-dimensional Edwards-Anderson 
model \cite{Bouc01,Bert02}. 
Although the typical length scales are merely of the 
order of a few lattice sizes, see e.g. \cite{Yosh02}, the relaxation times are
sufficiently large for a dynamical scaling to set in. 
While the expression (\ref{gl:tL}), if correct, 
points towards a cross-over behaviour between a simple power-law
scaling and an exponential scaling for $T<T_c$ in spin glasses
as would follow from the droplet model, 
it also suggests that at criticality simple power-law scaling 
(in the sense discussed above in section~3) should prevail.

{}From now on we consider the three- and the four-dimensional
Ising spin glass quenched to $T=T_c$ 
from a fully disordered initial state 
and discuss the dynamical scaling behaviour of critical two-time quantities.
We consider two classes of observables: (a) thermoremanent magnetization where
$h$ is constant up to a single jump and (b) alternating susceptibility, where
$h=h(t)$ is oscillatory. 

\begin{figure}[t]
\centerline{\epsfxsize=3.9in\epsfbox
{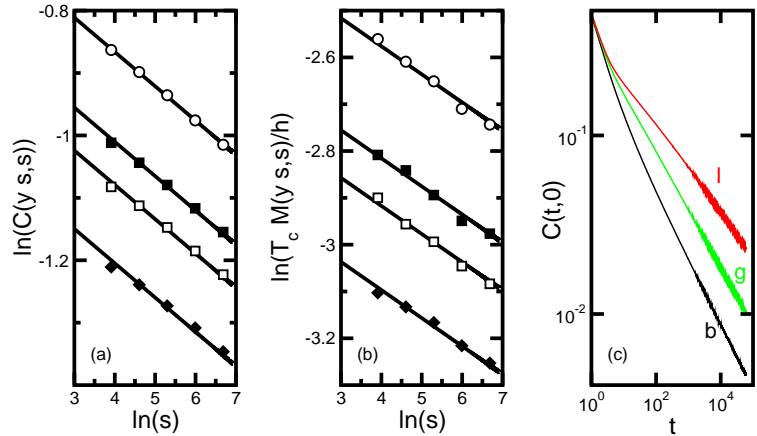}
}
\caption{Scaling of (a) the autocorrelation function $C(y s,s)$ and
(b) the thermoremanent magnetization $M(y s,s)$ of the critical 
three-dimensional Edwards-Anderson spin glass with a bimodal distribution of 
the couplings for (from top to bottom) $y= 5, 8, 10, 15$. The full curves are
$C(y s,s) = c_0 s^{-b}$ and $M(y s,s) = m_0 s^{-a}$, where $c_0,m_0$ were
fitted to the numerical data. 
(c) Temporal evolution of the
autocorrelation $C(t,0)$ in the three-dimensional cases (b: bimodal,
g: Gaussian, l: Laplacian). 
\label{Bild1}}
\end{figure}

\subsection{Thermoremanent magnetization} 

Let us first verify that the standard scaling forms (\ref{gl:sC}) and 
(\ref{gl:sR}) for the autocorrelation and the autoresponse are indeed
observed in {\em critical} spin glasses. From the data shown in 
figure~\ref{Bild1}
for the case of a bimodal distribution of the couplings in three dimensions
we conclude that both the correlation (figure~\ref{Bild1}a)
and integrated response (figure~\ref{Bild1}b) are
consistent with a power-law scaling. Similar results are obtained 
\cite{Henk05b,Plei05b}
for the two other distributions, both in three and four space dimensions.

{}From the slopes of figure~\ref{Bild1} we obtain the estimates for the 
exponents $a$ and $b$ gathered in table~\ref{tab1}.
Interestingly, 
we observe that $a=b$ within our numerical errors for a fixed
choice of the distribution of the couplings, but that the results for  
different distributions are different.
Even more intriguingly, other critical non-equilibrium dynamical quantities, 
as for example the
exponent $\lambda_C/z$ describing the decay of the correlations in the long-time
limit or the limit value of the fluctuation-dissipation ratio 
(\ref{gl:rfd}),  also vary strongly and systematically with the form of the
interaction distribution, see table~\ref{tab1} and figure~\ref{Bild1}c
\cite{Plei05b}. Here we emphasize the diagnostic usefulness 
of the {\em universal} limit
fluctuation-dissipation ratio $X_{\infty}$ whose numerical estimates vary
up to a factor of 2 between the various distributions. 
Having excluded the most probable sources of systematic errors \cite{Plei05b},
we interpret the numerical data as
strong evidence that in spin glasses critical out-of-equilibrium
quantities do depend on the exact form of the
distribution of the couplings. \footnote{
We recall that a similar result was
already seen for the disordered, but unfrustrated, Ising model in section~3.}
It is not yet clear whether this intriguing
property is only a purely dynamical effect.
Whereas in the past some studies of static critical quantities observed
a similar dependence on the distribution of the couplings \cite{Bern97}, recent studies
yielded for the different distributions values which agree
within error bars \cite{Jor06,Katz06}. 

\begin{table}[t]
\caption{
Nonequilibrium quantities of the critical 
Ising spin glass for bimodal, Gaussian, and Laplacian distributions of the 
nearest-neighbour couplings in $d=3$ and $d=4$ dimensions.
\label{tab1}}
\begin{center}
\begin{tabular}{|c|cccc|}  \hline
    & \multicolumn{4}{|c|}{bimodal}                     \\ \hline
$d$ & $a$      &      $b$  & $\lambda_C/z$ & $X_\infty$ \\ \hline
3   & 0.060(4) & 0.056(3)  & 0.362(5)~     &  0.12(1) ~ \\
4   & 0.18(1)~ & 0.180(5)~ & 0.615(10)     &  0.20(1)~  \\ \hline
 & \multicolumn{4}{|c|}{Gaussian} \\ \hline
$d$ & $a$      & $b$       & $\lambda_C/z$ &  $X_\infty$ \\ \hline
3   & 0.044(1) & 0.043(1)  & 0.320(5)      & 0.09(1)~ \\
4   & 0.169(4)~ & 0.171(2)~ & 0.58(1)~  & 0.175(10) \\ \hline
& \multicolumn{4}{|c|}{Laplacian} \\ \hline
$d$ & $a$      & $b$       & $\lambda_C/z$ &  $X_\infty$ \\ \hline
3   & 0.033(3) & 0.032(2)  & 0.259(2)      & 0.055(2) \\
4   & 0.143(5) & 0.140(3)  & 0.54(1)       & 0.13(1) \\ \hline

\end{tabular}\end{center}
\end{table}

\begin{figure}[tb]
\centerline{\psfig{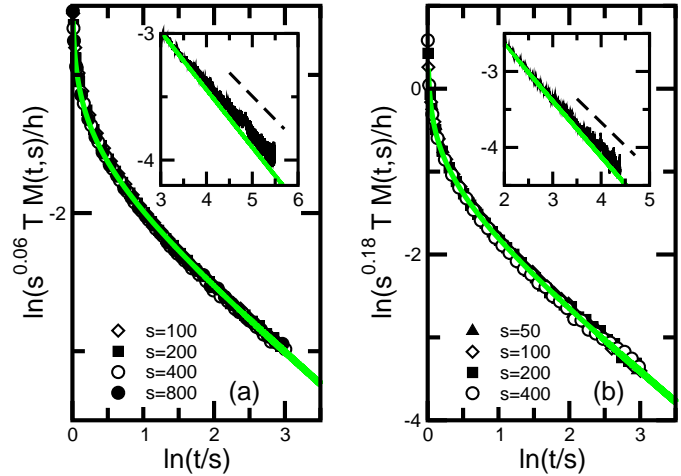}}
\caption{Scaling of the thermoremanent magnetization in the
Edwards-Anderson spin glass at criticality with 
a bimodal distribution of the couplings in (a) three
and (b) four dimensions. The full curve is the prediction
(\ref{gl:LSI}) of local scale-invariance. 
The results of very longs runs for a single waiting time
are shown in the insets where (a) $s=100$ in three dimenions
and (b) $s=25$ in four dimensions. 
\label{Bild3}} 
\end{figure}

Let us now turn to the scaling functions themselves. Plotting the
rescaled autocorrelator $s^b C(t,s)$ versus the scaling variable $t/s$ yields
a nice data collapse compatible with
a simple power-law scaling $L(t)\sim t^{1/z}$ \cite{Henk05b,Plei05b}.
In figure~\ref{Bild3} we show the two-time scaling of the
integrated response where again a collapse of the data
in terms of a simple power-law scaling is observed. The fact that both the 
autocorrelator
and the thermoremanent magnetization can be described in terms of such a
power-law scaling is evidence in favour of the time-dependent length-scale
(\ref{gl:tL}). 
In principle, one would like to extract an 
exponent $\lambda'_R/z$ from the slopes of the thermoremanent magnetization. 
It turns out that the values of $\lambda'_R/z$ thus obtained 
are significantly different from the
ones found for $\lambda_C/z$. For example, for the Ising spin glass with a 
bimodal distribution of the couplings one finds $\lambda'_R/z=0.45$ 
(resp. 0.72) in three resp. four dimensions. On the other hand, since
$X_{\infty}$ is finite, one must have $\lambda_R=\lambda_C$, but it may be 
necessary to go to very large values of $y=t/s$ in order to see this. 
Indeed, if one considers larger values
of $y=t/s$ as is shown in the insets of figure~\ref{Bild3}ab, one observes a 
passage from an effective exponent $\lambda_R'/z$ 
at intermediate values of the scaling variable $y$ to the truly
asymptotic value $\lambda_R/z$ at larger values of $y$. 
Because of this passage from
an effective exponent $\lambda'_R/z$ at intermediate values of $y=t/s$ to the
truly asymptotic value for larger $y$ we cannot expect the LSI
equation (\ref{gl:LSI}) to describe the scaling function $f_M(y)$ 
for all values of $y=t/s$.
In figure~\ref{Bild3}ab we compare the numerical data, 
in both three and four dimensions with the prediction (\ref{gl:LSI}) 
where we have
inserted the values of the exponents $a$ and $\lambda'_R/z$.
We find a nice agreement of the prediction (\ref{gl:LSI}) of local
scale-invariance with our data for $y=t/s$ not too large, but for very large
arguments the precise behaviour of the scaling function $f_M(y)$ cannot
be fully reproduced.

\subsection{Alternating susceptibility}

An interesting possibility to further test the behaviour of critical 
spin-glass models consists of working with a time-dependent (oscillating) 
magnetic field and to study simultaneously the dependence on time
and on the imposed oscillation angular frequency $\omega$. For a harmonic
magnetic field, it is common to consider the real and the imaginary part
of the magnetic susceptibility
\BEA
\chi'(\omega,t) = \int_0^t \!\D u\: R(t,u) \cos\left( \omega(t-u)\right)
\nonumber \\
\chi''(\omega,t) = \int_0^t \!\D u\: R(t,u) \sin\left( \omega(t-u)\right)
\label{gl:sus}
\EEA
where $R(t,s)$ is the linear response discussed above. In this setting, 
$1/\omega$ provides the second time-scale, the natural scaling variable
is $y=\omega t$ and the scaling regime should be reached in the
limit $\omega\to 0$ and $t\to\infty$. In many experiments and simulations,
one averages over at least one period of the oscillating field, 
see e.g. \cite{Picc01}. Then in a great variety
of glass-forming substances quenched to below or near to 
their glass transition point
one observes good but not always perfect evidence for an
$\omega t$-scaling behaviour  
of the following form of the period-averaged dissipative
(imaginary) part \cite{Joen02,Suzu03}
\BEQ \label{gl:chi2p}
\overline{\chi''}(\omega,t) = 
\chi''_{\rm st}(\omega) + \chi''_{\rm age}(\omega,t)
\;\; , \;\; 
\chi''_{\rm age}(\omega,t) \simeq A''_{\rm age} \left( \omega t\right)^{-b''}
\EEQ
where $\chi''_{\rm st}$ is thought of as a `stationary' contribution while
the ageing behaviour is described by $\chi''_{\rm age}$. The amplitude
$A''_{\rm age}$ and the exponent $b''$ are obtained from fits to the
experimental data. 
Similar scaling forms have been proposed for the dispersive (real) 
part $\overline{\chi'}$ but in practice the imaginary part is usually easier to 
measure. It is usually thought that $b'=b''$. 

Assuming the validity of LSI, we have derived the 
scaling relation \cite{Henk05a}
\BEQ \label{gl:bbaa}
b' = b'' = a - a'
\EEQ
where $a'$ is the exponent defined in eq.~(\ref{gl:Rf}). Hence this method
gives direct access to the exponent $a-a'$ and should be useful for 
future direct tests of whether $a$ and $a'$ are really different.

In order to derive (\ref{gl:bbaa}), we notice the central r\^ole 
played by the time difference $\tau=t-u$ in (\ref{gl:chi2p}) \cite{Henk02a}. 
Depending on its 
value either an equilibrium behaviour or else an ageing behaviour is obtained.
Recall from section~3 that there is a time-scale
$t_p\sim t^{\zeta}$ with $0<\zeta<1$ on which the transition between the
two regimes occurs \cite{Zipp00} such that $R(t,s)\simeq R_{\rm eq}(t-s)$ for 
$t-s\lesssim t_p$ and $R(t,s)=s^{-1-a}f_R(t/s)$ as given in eq.~(\ref{gl:Rf}) 
for $t-s\gtrsim t_p$. 
In addition, one measures for $u\approx t$ the response
with respect to a change in the initial conditions and then instead of
(\ref{gl:Rf}) one expects $R\approx R_{\rm ini}(t)\sim t^{-\lambda_R/z}$ 
\cite{Bray94}.\footnote{A similar argument is used in the derivation 
of eqs.~(\ref{gl:M},\ref{gl:LSI}) \cite{Henk02a,Henk03e}}.  
We therefore must introduce a 
further time-scale $t_{\eps}$ 
such that $t-t_{\eps}={\rm O}(1)$. Splitting the
integral into three terms corresponding to these three regimes, one has
\BEA
\lefteqn{ \chi''(\omega,t) = \int_0^t \!\D\tau\: R(t,t-\tau) \sin \omega\tau}
\nonumber \\
&=& \int_0^{t_p} \!\D\tau\: R(t,t-\tau) \sin \omega\tau +
\int_{t_p}^{t_{\eps}} \!\D\tau\: R(t,t-\tau) \sin \omega\tau 
\nonumber \\
& & +\int_{t_{\eps}}^t \!\D\tau\: R(t,t-\tau) \sin \omega\tau 
\nonumber \\
&\simeq& \int_{0}^{t_p} \!\D\tau\: R_{\rm eq}(\tau) \sin\omega\tau +
t^{-a}\int_{t_p/t}^{t_{\eps}/t} \!\D v\: f_R\left(\frac{1}{1-v}\right)
\frac{\sin \omega tv}{(1-v)^{1+a}}
\nonumber \\ 
& & +t^{-\lambda_R/z} \int_{t_{\eps}}^{t} \!\D\tau\: c_0 \sin\omega\tau
\nonumber \\
&=& \chi_1''(\omega) +
t^{-a}\int_{0}^{1} \!\D v\: f_R\left(\frac{1}{1-v}\right)
\frac{\sin \omega tv}{(1-v)^{1+a}} +
{\rm O}\left(t^{-\lambda_R/z}\right) \nonumber \\
&=& \chi_1''(\omega) + t^{-a} \chi_2''(\omega t) +
{\rm O}\left(t^{-\lambda_R/z}\right) \label{chipp} 
\EEA
In the third line, we used the asymptotic forms of $R(t,s)$ as described
above. This means that the cross-over between the equilibrium and the ageing 
regimes is assumed to be very rapid. While this is certainly correct for 
simple magnets, its validity in spin-glasses is far from obvious. 
In the last line, we restricted 
ourselves to the long-time
limit $t\to\infty$. Here, the function $\chi_1''(\omega)$ merely depends on the
equilibrium form of the linear response $R_{\rm eq}(t,s)$. It also 
becomes clear that the often-found stationary term 
in the {\em integrated} response 
\cite{Cate00,Coll00,Cugl02,Dupu04,Joen02,Kawa04,Rhei04,Rodr03,Suzu03} 
does not necessarily require the separation of a similar `stationary' part 
in the response function $R(t,s)$ itself. An analogous expression can be
derived for $\chi'(\omega,t)$.

We now insert the prediction (\ref{gl:Rf}) of LSI. Then the 
scaling functions read, together with their leading behaviour 
as $y\to\infty$ \cite{Henk05a} 
\BEA
\chi_2''(y) &=& 
f_0 B\left(1-a',\frac{\lambda_R}{z}-a\right) \,y^{1-a} 
\nonumber \\
& &  \times
{_2F_3}\left(\frac{1-a'}{2},\frac{2-a'}{2};\frac{3}{2},
\frac{1-a-a'}{2}+\frac{\lambda_R}{2z},\frac{2-a-a'}{2}+\frac{\lambda_R}{2z};
-\frac{y^2}{4}\right) 
\label{3:gl:cc} \\
&\simeq&  
f_1''  y^{a'-a} + f_2'' 
y^{-\lambda_R/z}\sin\left(y+\frac{\pi}{2}\left[a-\lambda_R/z\right]\right)
\nonumber \\
\chi_2'(y) &=&  
f_0 B\left(-a',\frac{\lambda_R}{z}-a\right) \,y^{-a}
\nonumber \\
& & \times
{_2F_3}\left(\frac{-a'}{2},\frac{1-a'}{2};\frac{1}{2},
\frac{-a-a'}{2}+\frac{\lambda_R}{2z},\frac{1-a-a'}{2}+\frac{\lambda_R}{2z};
-\frac{y^2}{4}\right)
\label{3:gl:ccc}\\
&\simeq&  
f_1' y^{a'-a} + f_2'
y^{-\lambda_R/z}\cos\left(y+\frac{\pi}{2}\left[a-\lambda_R/z\right]\right)
\nonumber 
\EEA
where ${}_2F_3$ is a hypergeometric function and $f_{1,2}'$ and $f_{1,2}''$ 
are known constants proportional to the normalization constant $f_0$. 
We see that there appear terms which decrease monotonously with $y$ 
but that there are
also oscillating terms. They are described by different exponents and must
be extracted by a different experimental setup. The oscillating terms follow
the oscillations of the external field and the decrease of the oscillation
amplitude gives a direct access to the exponent $\lambda_R/z$. 
On the other hand, in many experiments the data are averaged over one or several
periods of the external field. For $y$ sufficiently large, the contribution of
the oscillating term in eqs.~(\ref{3:gl:cc},\ref{3:gl:ccc}) vanishes after
averaging and then only
a simple algebraic component remains, which permits to extract the exponent
$a-a'$. For period-averaged data or else if $\lambda_R/z\geq a-a'$, the
leading behaviour for large arguments is
\BEQ
\chi_2'(y) \sim \chi_2''(y) \sim y^{a'-a}
\EEQ
and the scaling relations (\ref{gl:bbaa}) follow. Since the new exponent $a'$
enters in (\ref{gl:bbaa}), this suggests that no scaling relation between
$b'=b''$ and the other ageing exponents, with $a'$ excluded, exists. 

We now describe a test \cite{Henk05a} 
of the predictions eqs.~(\ref{3:gl:cc},\ref{3:gl:ccc}) in the 
three-dimensional Ising spin glass with bimodal disorder, quenched to its critical 
point $T_c\approx 1.19$. 
In order to study the alternating susceptibility far from equilibrium
we prepared the system in an uncorrelated initial state (corresponding
to an infinite initial temperature) before quenching it to the 
critical point at time $t=0$. At the same time an external oscillating
magnetic field 
\BEQ \label{gl:h}
h(t) = h_0 \cos \omega t
\EEQ
was switched on, with its amplitude fixed at $h_0=0.05$ which is well
inside the linear-response regime. We consider
different values of the angular frequency $\omega = 2 \pi/p$
with $p$ ranging from $50$ to $1600$. 
Numerically, the in-phase and the out-of-phase susceptibilities
are given by the expressions \cite{Ande96}
\BEQ
\chi''(\omega,t) =  m(t) \sin \omega t \;\; , \;\; 
\chi'(\omega,t) =  m(t) \cos \omega t  
\EEQ
with $m(t) = \sum\limits_{\vec{i}} \sigma_{\vec{i}}(t)$ but the
equilibrium parts of $\chi''$ and $\chi'$ must be subtracted off \cite{Henk05a}.

\begin{figure}
\centerline{\epsfxsize=4.0in\epsfbox
{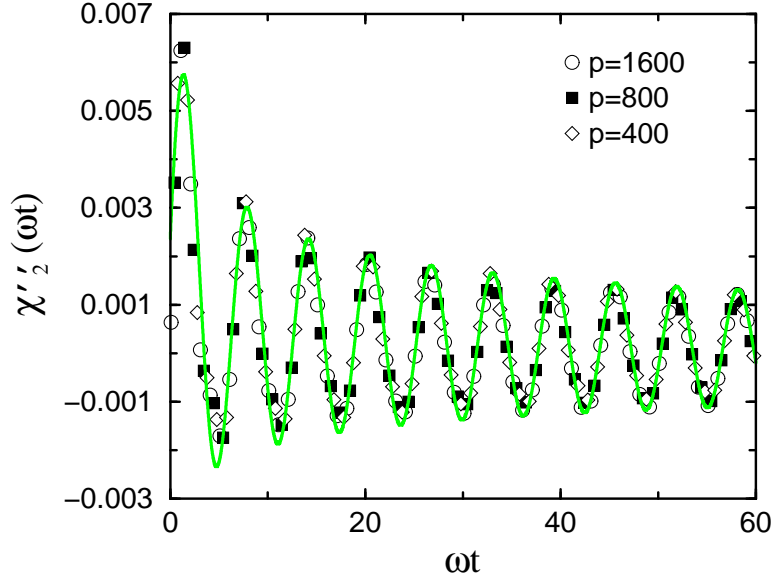}}
\caption{Scaling of the dissipative part $\chi_2''(\omega t)$ 
of the alternating susceptibility as function of the 
scaling variable $\omega t$ for different angular
frequencies $\omega= 2 \pi/p$ with $p= 1600$, 800, and 400. The full curve
is the theoretical prediction (\ref{3:gl:cc}) with 
$f_0 = 0.002$ and $a' = -0.70$ but which has also been shifted 
horizontally by $y\to y+\Delta y$, with $\Delta y= -0.45$, see text. 
Statistical error bars are smaller than the symbol sizes.
\label{fig1}}
\end{figure}

\begin{figure}
\centerline{\epsfxsize=4.0in\epsfbox
{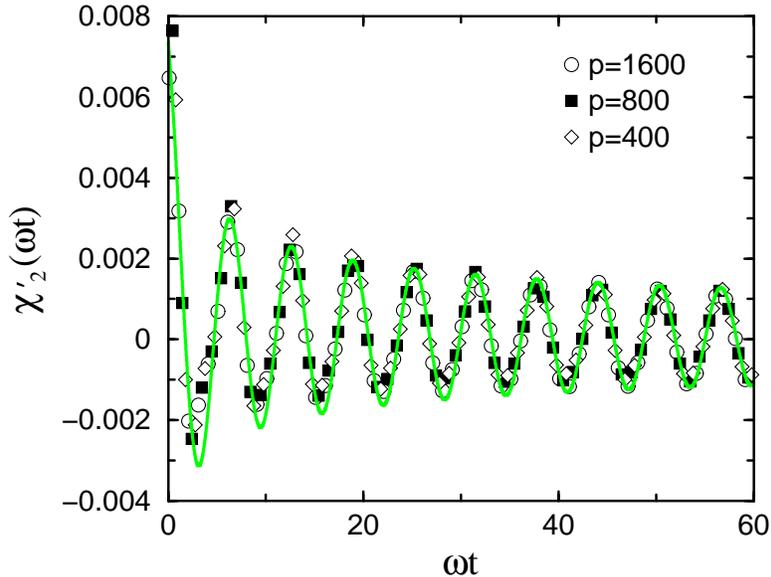}}
\caption{The same as in Figure \ref{fig1}, but now for the dispersive part 
$\chi_2'(\omega t)$. The full curve is the shifted theoretical prediction 
(\ref{3:gl:ccc}) with $f_0 = 0.002$ and $a' = -0.70$.
\label{fig2}}
\end{figure}

In figures~\ref{fig1} and \ref{fig2} \cite{Henk05a} we compare the expected
scaling behaviour of the ageing part
\BEQ
\chi_2'' = \chi_2''(\omega t) ~~~ \mbox{and} ~~~
\chi_2' = \chi_2'(\omega t).
\EEQ
For the larger values of $p$, corresponding to the smaller values
of $\omega$, one observes a very good data collapse for both quantities, which
furnishes clear evidence in favour of a power-law scaling at $T=T_c$. 
The value $a=0.060(4)$ determined
previously from the decay of the
thermoremanent magnetization \cite{Henk05b,Plei05b} was used. 
For smaller values of $p$, the collapse is less good. It is possible 
that for the corresponding values
of $\omega$ the dynamical scaling regime is not yet reached.

The data shown in these two figures can in principle be compared directly
with the analytical predictions (\ref{3:gl:cc}) and (\ref{3:gl:ccc}),
{\em provided} however, that the positions 
of maxima of $\chi_2''$ and
$\chi_2'$ are shifted by the amount $\Delta y\approx -0.45$ when compared with
the positions obtained in the analytical treatment of the previous section.
Where this shift comes from is not understood. 
It is possible, however, that the cross-over
between the equilibrium and the ageing regimes is not almost instantaneous
in contrast to what we assumed in the derivation of eq.~(\ref{chipp}). 

Since both $a$ and $\lambda_R/z$ were known before \cite{Henk05b,Plei05b} 
the only free parameters in this comparison are the amplitude $f_0$
and the exponent $a'$. The final estimates are \cite{Henk05a}
\BEQ \label{3:gl:af}
a' = -0.70(3) \;\; , \;\; f_0 = 0.00203(1) 
\EEQ
and the fit is compared to the data in the figures. 
These values (\ref{3:gl:af}) of the parameters describe consistently 
both $\chi_2''$ and $\chi_2'$.
Although some discrepancies are observed for small values of $y=\omega t$,
the overall agreement between the simulation and the shifted theoretical
prediction is very good. Clearly, LSI captures an essential part of the
ageing behaviour of critical spin glass models. 


We finish with a discussion of existing experimental results on the scaling
of the alternating susceptibility. 
A detailed discussion of a possible scaling of the alternating susceptibility
was presented by Suzuki and Suzuki for the short-ranged Ising spin glass
Cu$_{0.5}$Co$_{0.5}$Cl$_2$-FeCl$_3$ graphite bi-intercalation 
compound \cite{Suzu03}. After a rapid quench to below the glass temperature
$T_g=3.92(11) {\rm K}$, they measure $\chi''(\omega,t)$ for fixed $\omega$ and
find that their (period-averaged) data are well-fitted by the power-law
\BEQ
\overline{\chi''}(\omega,t) = \chi_0''(\omega) + A''(\omega) t^{-b''}
\EEQ
While the fitted exponent $b''$ depends only slightly on $\omega$, they further
show evidence for a power-law $A''(\omega)=A_0'' \omega^{-\mu''}$ and find
that {\it ``\ldots the value of $\mu''$ is almost the same as that of 
$b''$''} \cite{Suzu03}.
In this way, they arrive at the $\omega t$-scaling form
\BEQ
\overline{\chi''}(\omega,t) = 
\chi_0''(\omega) + A_0'' \left( \omega t\right)^{-b''}
\EEQ
and a similar form for $\overline{\chi'}(\omega,t)$ 
where {\it ``\ldots $b'$ and $b''$ 
are of the same order at the same temperature''} \cite{Suzu03}.
Experimentally measured values of the exponent $b''$ (and also $b'$) of some
materials are collected in table~\ref{table1}. Comparing the
experimental results of Suzuki and Suzuki \cite{Suzu03} with the theoretical
scaling form (\ref{chipp}), one sees:
\begin{enumerate}
\item the experimental evidence for a pure
$\omega t$-scaling indicates that the exponent $a$ must indeed be very small. 
\item when considering the leading behaviour for $y=\omega t$ large 
(their data go up to $y\lesssim 10^6$ \cite{Suzu03}) and recalling that 
the experimental data are averaged over at least one period of the 
external field, one can read off  
\BEQ
b''= a - a' \;\; , \;\; \mu'' = -a'
\EEQ
and the observed \cite{Suzu03} near equality $b''\approx \mu''$ is 
again consistent with $a$ being negligibly small. 
\item the data
are consistent with the theoretically requested relation $b'=b''$. 
\end{enumerate}

\begin{table}
\caption[T1]{Measured values of the exponents $b''$ and $b'$ in several glassy
materials, using the scaling form \protect{(\ref{gl:chi2p})}. 
Here $T_g$ stands for the glass transition temperature and 
$T$ is the temperature where the data were taken. 
For Fe$_{0.5}$Mn$_{0.5}$TiO$_3$ and CdCr$_{1.7}$In$_{0.3}$S$_4$ the
relation $b'=b''$ was assumed. The simulational results in the critical
Ising spin glass are also included. \label{table1}}
\begin{tabular}{|ll|llll|ll|} \hline\hline
\multicolumn{2}{|c|}{Material} & \multicolumn{1}{c}{$T_g \mbox{\rm [K]}$} & 
\multicolumn{1}{c}{$T \mbox{\rm [K]}$} & \multicolumn{1}{c}{$b''$} & 
\multicolumn{1}{c|}{$b'$} & Ref. & \\ \hline
\multicolumn{2}{|l|}{Cu$_{0.5}$Co$_{0.5}$Cl$_2$-FeCl$_3$} & 
3.92(11) & 3.25 & 0.01(4) & 0.08(3) & \cite{Suzu03} & Ising\\
\multicolumn{2}{|l|}{-- GBIC} &     & 3.5  & 0.017(32) & 0.05(2) && spin glass\\
 & &     & 3.75 & 0.16(3)   & 0.20(2) & & \\
 & &     & 3.85 & 0.15(3)   &         & & \\
 & &     & 3.95 & 0.16(4)   & 0.20(2) & & \\ \hline 
\multicolumn{2}{|l|}{Fe$_{0.5}$Mn$_{0.5}$TiO$_3$} & 
20.7     & 15   & 0.14(3)   & & \cite{Dupu01,Dupu02} & Ising  \\
 & &     & 19   & 0.14(3)   &         & & spin glass \\ \hline
\multicolumn{2}{|l|}{CdCr$_{1.7}$In$_{0.3}$S$_4$} & 
16.7     & 12   & 0.18(3)   & & \cite{Dupu01,Dupu02} & Heisenberg  \\
 & &     & 14   & 0.18(3)   &         & & spin glass\\ \hline 
CdCr$_{2x}$In$_{2-2x}$S$_4$ & $x=0.95$ & 
70       & 8    & 0.2       & & \cite{Dupu02a,Dupu02} & disordered\\
 & &     & 67   & 0.2       &         & & ferromagnet \\
 & $x=0.90$ 
         & 50 & 42 & 0.20   &         & \cite{Vinc00,Dupu02} & \\ \hline 
\multicolumn{2}{|l|}{Pb(Mg$_{1/3}$Nb$_{2/3}$)O$_3$} & 
$\sim 220$ & $\lesssim 220$ & 0.17 &  & \cite{Coll00} & relaxor  
\\ 
 & & & & & & & ferroelectric \\ \hline\hline 
theory & ($T=T_c$) & & & 0.76(3) & 0.76(3) & \cite{Henk05a} & Ising \\
& & & & & & & spin glass \\ \hline\hline
\end{tabular}
\end{table}

Similar values of $b''$ were observed for several other materials, quite
independently of the precise physical nature as can be seen from 
table~\ref{table1}, but the errors are still too large to permit a discussion
of the universality of the exponents. 
However, the experimental data are in many of these
systems at least as well described by a logarithmic scaling as expected from
the droplet theory \cite{Dupu02,Joen02,Dupu01}. Furthermore, in several
systems also strong deviations from a simple $\omega t$-scaling were found,
see \cite{Coll01}. Finally, we mention that in systems like 
$\beta$-hydroquinol-clathrate \cite{Rhei04} or even simple liquids like
glycerol \cite{Lehe98} a power-law dependence of the form 
$\chi''_{\rm age}\sim t^{-a}$ or $\chi'_{\rm age}\sim t^{-a}$ was observed. 
All in all, it is at present not completely clear why one should find
a simple $\omega t$-scaling of the alternating susceptibility whereas
subageing is frequenlt admitted for the thermoremanent magnetization. 

A last important observation follows from comparing the experimental data
for the exponents $b'=b''\approx 0.1 - 0.2$ for all the $3D$ materials 
studied so far with the theoretical estimate $b''\simeq 0.76$ derived from the
$3D$ Ising spin glass with binary disorder. The values are very far from each 
other and this huge difference calls for an explanation. 
Could this be seen as an indication that the spin glass models considered
by theorists only capture imperfectly what is going on in real materials~?

\section{Discussion}

The available evidence for universality in the ageing behaviour seen in  either
glassy materials, disordered systems or else simple many-body systems without
disorder may be taken, in combination with dynamical scaling, as suggestive of
the existence of deeper dynamical symmetries in such systems. We have advocated
here the point of view that such a new symmetry might be sought through an
extension of standard dynamical scaling towards a local scale-invariance. 
The present formulation of the theory has been built around dynamical scaling
as it occurs in full ageing, but this was mainly for reasons of technical 
simplicity and one could consider different forms of dynamical scaling, with
a modified form of the Lie algebra generator $X_0$, if required. 

Even within this specific context, it is in principle possible to generalize in
many different ways. For that reason, it is important to be able to test
quantitatively some of the consequences of local scale-invariance. Here the
calculation of the two-time response function $R(t,s)$ is one of the easiest
tasks and has the advantage that prediction are relatively easy to test
in simulations, for example by considering an integrated response such as
the thermoremanent magnetization M$_{\rm TRM}(t,s)$. 

The central idea of LSI is the decomposition of the stochastic Langevin equation
into a {\bf deterministic part} which can possess local scale-invariance and
into a {\bf noise part} which does break this. Remarkably, the noise part is
usually built in such a way that any $n$-point function of the full theory can 
be reduced exactly to a certain $m$-point function of the noiseless theory 
(where $m\geq n$). This reduction depends on the Bargman superselection rules
which are proven only for $z=2$ at the time of writing. Then the symmetries of the
noiseless part yield conditions on the form of the $m$-point functions. In
certain cases, such as the two-time response functions, these constraints are
enough to fix the functional form, see eq.~(\ref{gl:Rf}). 

In models where the deterministic part of the Langevin equation is 
{\em linear}, LSI can be proven, 
for example in the spherical model \cite{Pico04},
the $1D$ Glauber-Ising model \cite{Pico04,Henk06a}, bosonic particle-reaction
models \cite{Baum05b} or certain simple growth models \cite{Roet06,Baum06e}. 
Furthermore, there is a systematic way to construct LSI-invariant 
non-linear equations \cite{Stoi05,Baum05b}. Since these are not identical to the
Langevin equations one usually starts from \cite{Hohe77}, numerical evidence
for those systems is crucial, in the absence of exact solutions.  

In order to be able to carry out meaningful tests, one must first show that
simple ageing (on which LSI is built) holds true. For non-conserved 
ferromagnets this is thought to be certain \cite{Bray94} and we have discussed
the evidence that this remains so for disordered, non-glassy Ising models.
In most numerically studied systems, only $M_{\rm TRM}(t,s)$ has been compared
with LSI. Provided that corrections to scaling are treated carefully, in all
cases the results are consistent with LSI, see tables~\ref{Tabelle1} and
\ref{Tabelle2}. The fact that $M_{\rm TRM}(t,s)$ as measured in a {\em noisy} 
system is consistent with the integral of (\ref{gl:Rf}) found from the
symmetries of the {\em noiseless} theory, even if $z\ne 2$, is a strong
indication that an extension of the Bargman superselection rules to $z\ne 2$
should exist. 

At present, the only other observables measured have been space-time
responses and autocorrelations in the two- and three-dimensional Ising models\cite{Henk03b,Henk04b}
and in the two-dimensional Potts-$q$ models ($q=2,3,8$) \cite{Lore06}. It would be important
to consider different quantities, e.g. alternating susceptibilities. 

When turning to critical spin glasses, although there is evidence that simple
ageing applies, the agreement of the data with LSI was only partial. It should 
be recalled that the analysis performed relied on a rapid change between
quasi-static and scaling behaviour, which might be an oversimplification. 
Remarkably, in the presence of disorder we have seen evidence that the
non-equilibrium exponents may depend on the distribution of the random exchange
couplings $J_{i,j}$, independently on whether the system was glassy or not. 
Since this is a highly controversial issue, further tests would 
be most welcome. 

At present, the available evidence\footnote{Of course models such as the 
celebrated zero-range process, see \cite{Evan05,Godr06}, which lacks a proper 
spatial structure should be excluded from this kind of consideration.} 
from both analytical solved models and numerical studies supports the predictions of local scale-invariance for the two-time response (and in some cases also correlation) functions. This suggests that the kind of Langevin equation used in the description of these models should possess a 
fundamental, hitherto unsuspected and non-trivial
dynamical symmetry underlying certain stochastic Langevin equations. 
Already the present examples show that the calculation of time-dependent
quantities such as the thermoremanent magnetization is considerably simplified and leads to results not available as yet by other methods. Local scale-invariance, similar in spirit to
the tremendously successful techniques of conformal field-theory, has the potential of leading to a much more profound understanding of non-equilibrium
critical phenomena. 

Finally, comparison of simulational results with experimental evidence for 
both glassy and non-glassy systems (see tables~\ref{tabl4} \& \ref{table1}) 
points to some systematic differences which need to be clarified. \\

\noindent {\bf Acknowledgements :} 
We thank Wolfhard Janke for his invitation to write this review,
F. Baumann, I.A. Campbell, S.B. Dutta, T. Enss, A. Gambassi, A. R\"othlein 
and J. Unterberger for fruitful collaborations on the th\'ematique 
reviewed here and acknowledge the support by the Deutsche Forschungsgemeinschaft
through grant no. PL 323/2.
This work was also supported by the franco-german binational
programme PROCOPE and by CINES Montpellier (projet pmn2095). 

\newpage 
  
%

\end{document}